\documentclass[12pt]{article}

\usepackage[utf8x]{inputenc}
\usepackage[left=2.5cm,right=2.5cm,top=2.5cm,bottom=3cm]{geometry}
\usepackage[english]{babel}
\usepackage{graphicx}
\usepackage{mathtools}
\usepackage{braket}
\usepackage{pifont}
\usepackage[dvipsnames]{xcolor}
\usepackage[hidelinks]{hyperref}
\usepackage{amssymb}
\usepackage{tikz}
\usepackage[compat=1.1.0]{tikz-feynman}
\usepackage{contour}
\usepackage{enumitem}
\usepackage{mathrsfs}
\usepackage{booktabs}
\usepackage{tabularx}
\usepackage{afterpage}
\usepackage{placeins}
\usepackage{array}
\usepackage{subcaption}
\usepackage{float}
\usepackage{dsfont}
\usepackage[final]{microtype}
\usepackage[
    backend=biber,
    style=numeric-comp,
    sorting=none,
    sortcites=true,  
    maxnames=3,      
    minnames=1,       
    isbn=false,      
    url=false,       
    doi=false,       
    eprint=true      
]{biblatex}
\usepackage{bm} 

\renewcommand{\thesection}{\arabic{section}}
\renewcommand{\thesubsection}{\thesection.\arabic{subsection}}

\makeatletter
\renewcommand{\p@subsection}{} 
\makeatother

\makeatletter
\@addtoreset{equation}{section}
\makeatother

\begin{document}

\begin{flushright}
YITP-26-42
\end{flushright}

\vspace{1cm}

\begin{center}
    {\Large \bf Gravitational Properties of the Monopole Bag} \\[0.5cm]
    
    {\large Yu Komiya$^1$ and Fumihiro Takayama$^1$} \\[0.2cm]
    {\it $^1$Yukawa Institute for Theoretical Physics, Kyoto University, Kyoto 606-8502, Japan} \\[0.2cm]
    {\it E-mail: yu.komiya@yukawa.kyoto-u.ac.jp, takayama@yukawa.kyoto-u.ac.jp} \\[0.5cm]
    
    April 2026
\end{center}

\vspace{0.5cm}

\begin{abstract}
Axionic cosmologies constitute a class of models with phenomenologically rich symmetry breaking in the early universe. In the case where monopoles are present in such a background, the axion profile may be deformed; it is possible to construct a ``monopole bag" state composed of a central monopole within a closed axion domain wall. We consider the gravitational properties of this composite defect, and propose that there exist both a horizonless and a black hole-like static solution for different input parameters. For the latter case, we present our results as a first step towards a new construction of a dyonic regular black hole, given the monopole bag's potential ability to evade the usual singular gravitational collapse and retain a non-trivial axionic profile by the exotic electromagnetic properties of an axionic Chern-Simons term. 
\end{abstract}

\vspace{1cm}

\newpage
\tableofcontents
\newpage


\pagestyle{plain}

\renewcommand{\thesection}{\arabic{section}}
\renewcommand{\thesubsection}{\arabic{section}.\arabic{subsection}}



\section{Introduction}
\label{sec: intro}

The early Universe is a period with extreme and yet mysterious conditions. Theories about the evolution, composition, and consequences of this epoch are diverse; with our rapidly advancing generations of both ground-based and space-faring probes offering a multifaceted set of tests, it has become increasingly important to pursue the formulation of robust models with physical implications that may shed light on some of the greatest unsolved problems in modern physics.

One of these open questions pertains to the exact inflationary mechanism (should there indeed have been one) that gave rise to the temperatures, structures, and particles making up our present Universe. From CMB measurements by PLANCK 2018 \cite{collaboration_planck_2020} to the latest BAO results by DESI \cite{collaboration_desi_2025} (which are thought to be the most precise measurements ever taken of a cosmological parameter), the number of viable theories has been constrained with increasing accuracy. However, present results are not entirely convincingly compatible with each other \cite{ferreira_bao-cmb_2026}. The range of proposals for each aspect of various primordial processes constitutes our efforts towards fully understanding the history of our Universe.

Another outstanding problem pertains to the nature of dark matter \cite{cirelli_dark_2025}. Unlike ordinary (baryonic) matter, this elusive species interacts exclusively gravitationally, therefore the validity of candidates is guided only by the constraints we can place on its mass and behaviour through cosmological and astrophysical observations. Examples of objects proposed to make up dark matter include new particles (e.g. axions) \cite{preskill_cosmology_1983, marsh_axion_2016}, early universe species such as primordial black holes (PBHs) \cite{carr_black_1974, green_primordial_2021}, and other theoretical structures such as gravitating monopoles \cite{breitenlohner_gravitating_1992, sato_unified_2018}.

The dark matter mystery and the successful large-scale structure formation expected from nearly scale-invariant quantum seeds are intimately connected, and it remains challenging to describe smaller-scale structures where non-linear growth of gravitational clustering becomes important. Some traces of new physics originating from the early Universe might still be hidden inside the small-scale structures. Dark matter constituents and the resulting small-scale objects (which might be particles, scalar condensates or other extended objects) could offer fresh possibilities for both exotic astrophysical evolution of stars and stellar clusters, and the roles of dark matter beyond its gravitational properties. Further theoretical understanding of their dynamics as well as stationary states may clarify how primordial information is stored. Relatedly, the connection between core properties of defects and UV processes is intricate. In particular, in monopole systems, one may expect that a deeper understanding of processes such as fermion scattering in the core region \cite{callan_dyon-fermion_1982, loladze_monopole-fermion_2024} and decay channels described by both standard model and beyond (e.g. GUT) \cite{sonoda_decay_1984, hook_dyonic_2025} may assist our understanding of both low-energy defect dynamics and microscopic properties of core regions.

Finally, there also exist astrophysical questions worth investigating in addition to the phenomenological and cosmological puzzles mentioned above. Black holes in particular are central to many research areas owing to their status as an object at the intersection between quantum and gravitational physics. Questions about the true structure of the centre of black holes, the formation of extreme objects and their detection, the evolution of objects with black hole exteriors via Hawking radiation, and the presence of such gravitating species at late times remain popular topics \cite{cardoso_testing_2019, chen_black_2015, ansoldi_spherical_2008}. 

The study we have undertaken lives at the interface between these branches of physics. We start from the Peccei-Quinn (PQ) mechanism \cite{peccei_cp_1977} in a QCD scenario as a phenomenologically-motivated introduction of the axion. These hypothetical particles are well-known to source rich early universe dynamics \cite{kim_light_1987}, and solve the strong CP problem through the PQ mechanism in the case of QCD. Closely related are pseudo-Nambu-Goldstone bosons originating from a more general approximate $U(1)$ symmetry; these scenarios may be collectively referred to as axion-like particle (ALP) models. Notably, a spectrum of light, axion-like states can arise naturally from string compactifications \cite{arvanitaki_string_2010, conlon_qcd_2006, svrcek_axions_2006}, where the QCD axion is one realisation encompassed by this prediction. From such frameworks, one may derive classes of axion cosmologies integrated into a cohesive description of early times in which successive symmetry breaking gives rise to topological defects around the inflationary epoch \cite{kibble_topology_1976, hiramatsu_production_2012}. These defects result from a non-trivial vacuum manifold, where we focus on axion domain walls in particular as two-dimensional defects separating regions of degenerate vacua formed during the spontaneous breaking of a discrete symmetry. Then, we propose the presence of magnetic monopoles in addition to axion domain walls, which we highlight as an interesting combination of defects since their unique interaction is known to give rise to non-singular stable states in the case of a monopole enclosed within a spherical axion domain wall. This state was first studied and christened the ``monopole bag" by Kogan \cite{kogan_kaluza-klein_1993}. We investigate the gravitating properties of such a system for the first time, which then led us to consider the possible relevance of our scenario to the construction of a magnetic (or ``dyonic", if electric charge is also present\footnote{While ``dyon" strictly refers to an electrically charged monopole, it is often used interchangeably with ``monopole" in the literature. The reader may also notice instances in this work where both terms are valid within a statement given suitable adjustments to the nuance.}) regular black hole. This is a class of objects that has been theorised to exist and built via a range of different methods. Primarily, the gravitational requirements present an exciting challenge promising to enrich neighbouring fields of study, such as that of ``gravastars" \cite{beltracchi_slowly_2022, uchikata_slowly_2016, mazur_gravitational_2004}, exotically interacting systems \cite{felice_exotic_2025, bizon_gravitating_1994}, and regularised black holes \cite{mbonye_nonsingular_2005, bronnikov_regular_2001, bokulic_conundrum_2025}. Furthermore, both defects such as monopoles and stable black hole-type objects have been considered possible components of dark matter. The conditions surrounding their formation and the observational bounds on their present abundance form an important tie between primordial, high-energy theories and late-time confirmation via testable parameters.

The structure of this paper can be outlined as follows. In §\ref{sec: adw}, we review the underlying rationale for our scenario, namely the theory associated with axion models. This leads to a discussion of the interaction between axions and monopoles forming the monopole bag in §\ref{sec: mbag1}, and an analysis of the analogous phenomena in curved spacetime in §\ref{sec: amcurve}. The latter forms the crux of our investigation, the physical consequences of which will be discussed in §\ref{sec: physth}. Finally, we summarise our findings and consider future directions in §\ref{sec: sum}.


\section{Axion Domain Walls}
\label{sec: adw}

We will begin by exemplifying a way in which axion domain walls may arise by largely adhering to well-studied, standard arguments presented by the PQ mechanism in the context of QCD. Importantly, this method of producing domain walls is not unique, therefore we will attempt to make clear the branches in the formation pathway by highlighting where generalisations to ALPs are possible. As much of our upcoming analysis will prove deeply model-dependent, it is beneficial for maintaining consistency and clarity to begin with a concrete theoretical model in mind while not sacrificing awareness of alternatives that may offer a broader perspective. Therefore, the reader should note that future references to ``axions" and ``axion domain walls" are not limited to QCD theories, but instead may equally be substituted for statements about ALPs and ALP domain walls unless one or the other case is expressly under discussion. Throughout this work, we will adopt natural units in which $\hbar=c=1$, and the signature $(- + + \,+)$.


\subsection{Low-Energy Effective Theory for Axions}
\label{sec: eft}

The PQ mechanism is a popular solution to the famous strong CP problem in particle physics \cite{peccei_cp_1977}. The essence of the theory is to promote the phase of a PQ field $\varphi$ to a dynamical variable, permitting cancellation with the strong CP angle in a natural manner and thus obeying the vanishingly small value dictated by the lack of violation observed in experiments.

The PQ field $\varphi$ is a dynamical complex scalar field spontaneously breaking a global $U(1)_{\text{PQ}}$ symmetry at energy scale $v$; a typical Lagrangian takes the form
\begin{equation}
    \mathcal{L} = -\frac{1}{2}|\partial_\mu\varphi|^2 - \frac{\lambda}{4}\left( |\varphi|^2 - v^2 \right)^2,
\end{equation}
such that $\varphi$ has the vacuum expectation value
\begin{eqnarray}
    \varphi = ve^{i\vartheta} = ve^{ia/f_a}
\end{eqnarray}
with $f_a = v/N_{\text{DW}}$ being the so-called axion decay constant and $N_{\text{DW}}$ is known as the domain wall number. The Nambu-Goldstone boson $a$ of the broken $U(1)_{\text{PQ}}$ symmetry is known as the axion, which is often conveniently written in its dimensionless form $\vartheta$. Several alternative axion models with a more general $U(1)$ global symmetry have also been proposed: known as ALP models, they may be from different origins and motivations \cite{dunsky_primordial_2024, peccei_strong_2008, hiramatsu_axion_2013, gelmini_primordial_2023, vilenkin_cosmic_2001, vilenkin_cosmic_1982}. The $\mathbb{Z}_{N_{\text{DW}}}$-symmetric effective Lagrangian density for the axion at low-energy is thus
\begin{equation}
\label{eq: LEFT}
    \mathcal{L} = -\frac{1}{2}f_a^2\partial_\mu\vartheta\partial^\mu\vartheta - V(\vartheta),\text{ } V(\vartheta)=\frac{m_a^2f_a^2}{N_\text{DW}^2}\left[ 1-\cos\left( N_\text{DW}\vartheta \right) \right],
\end{equation}
in which $m_a$ is axion mass. In the example of QCD, the potential $V(\vartheta)$ and non-trivial axion mass arises from non-perturbative QCD topological fluctuations (often described as instanton effects), which become significant after the QCD crossover as confinement develops \cite{borsanyi_lattice_2016}. More generally, in ALP models where there is a $U(1)$ anomaly under a strongly coupled theory, such a potential can be induced as temperatures approach confinement scale \cite{marsh_axion_2016, cortona_qcd_2016}. It is evident from the form of the potential that vacua occur at $\vartheta  = 2\pi n$, $n\in\mathbb{Z}$, and $\vartheta$ is invariant under transformation by $2\pi$. Separating regions with different discrete vacua are axion domain walls.


\subsection{The Cosmology of Axion Domain Walls}
\label{sec: setup}
We start by considering a well-studied scenario in which PQ symmetry breaking and the QCD crossover (or corresponding strong confinement transition for the ALP model) both occur sometime after inflation; a detailed description of these processes can be found in, for instance, \cite{marsh_axion_2016, sikivie_axion_2008}. Initial PQ symmetry breaking occurs at temperatures $T<v$, forming axion strings as the Goldstone bosons $\vartheta$ wind from $0$ to $2\pi$. PQ symmetry is further explicitly broken to $\mathbb{Z}_{N_{\text{DW}}}$ once temperatures cool sufficiently for non-perturbative effects to activate; they generate the characteristic low-energy potential $V(\vartheta)$ above in (\ref{eq: LEFT}). Crucially, we noted that this potential has a series of degenerate vacua at $\vartheta=2\pi n/N_{\text{DW}}\text{ }(n\in\mathbb{Z}_{N_{\text{DW}}})$, resulting in domain wall formation at the interfaces between different $\vartheta$-valued vacua. These domain walls form with axion strings as their boundaries and may be thought of as solitonic configurations of the axion field, evolving as surfaces possessing energy density and reacting to pressure \cite{hiramatsu_evolution_2011}. For $N_{\text{DW}}>1$, $N_{\text{DW}}$ walls are connected to each string boundary such that the string-wall network is stable, and would therefore give rise to a so-called ``domain wall problem" \cite{dine_remarks_2023}. This entails the scenario where walls over-dominate due to their energy density falling as $a^{-1}$ with the scale factor $a$ (which is much slower than radiation and matter both) and thus provoking a deviation from our standard cosmological model. The problem may be circumvented via additional higher order terms (such as a bias potential) that break the degeneracy of the vacua and lead to the annihilation of the network \cite{gelmini_cosmology_1989}. Alternatively, the network is unstable and hence cosmically safe if $N_{\text{DW}}=1$, which we will assume for our analysis henceforth.\footnote{See for instance \cite{lazarides_axion_1982} for motivation behind such a selection.} Of particular interest to us are closed walls, which can be generated either at formation, or through interactions near annihilation time \cite{gelmini_primordial_2023}. Thus, we may ignore the dynamics of the wall-string network from now on and instead consider these roughly spherically symmetric, closed domain walls as isolated dynamical objects. For the same reason, our choice $N_{\text{DW}}=1$ serves only to simplify ensuing analysis with a physically motivated and well-studied choice; the addition of a bias potential, for instance, to retain compatibility with standard cosmology at $N_{\text{DW}}>1$ would not compromise the setup with which the remainder of this paper will be concerned.


\subsection{Electromagnetic Properties}

While domain walls do not intrinsically generate electromagnetic (EM) fields, they display unique EM properties when placed in an EM background. These behaviours originate from the coupling between the axion field and EM fields \cite{huang_structure_1985},
\begin{equation}
    \frac{\alpha N_{\text{DW}}\vartheta}{8\pi}F_{\mu\nu}\tilde{F}^{\mu\nu},
\end{equation}
in which $\alpha=e^2/4\pi$ is the fine structure constant, and $\tilde{F}^{\mu\nu} = \epsilon^{\mu\nu\rho\sigma}F_{\rho\sigma}/2$ is the dual field strength tensor. While usually such a term is a physically irrelevant total derivative, the dynamical $\vartheta$ and the instanton effects at strong coupling mean it cannot be trivially dropped. In the language of electric ($\vec{E}$) and magnetic ($\vec{B}$) fields, this term corresponds to a coupling
\begin{equation}
    \sim \vartheta\vec{E}\cdot\vec{B},
\end{equation}
and the overall EM Lagrangian of such an axion model is \cite{huang_structure_1985, sato_unified_2018}
\begin{equation}
\label{eq: EMLagr}
    \mathcal{L}_{EM} = -\frac{1}{2}f_a^2\partial_\mu\vartheta\partial^\mu\vartheta - \frac{m_a^2f_a^2}{N_{\text{DW}}^2}\left[ 1-\cos\left( N_{\text{DW}}\vartheta \right) \right] -\frac{1}{4}F_{\mu\nu}F^{\mu\nu} + \frac{\alpha N_{\text{DW}}\vartheta}{8\pi}F_{\mu\nu}\tilde{F}^{\mu\nu}.
\end{equation} 
While we are observing axion monodromy at work---given that the EM terms do not seem to recover a periodic form but instead take a quartic form in our low-energy estimate---the model must possess some UV completion in which higher order effects such as non-linear terms of quantum origin may well permit periodicity preservation \cite{silverstein_monodromy_2008, agrawal_monodromic_2024}.\footnote{Note, however, that the period does not necessarily coincide with that of the original axion potential. Additionally, a non-linear axion coupling with the EM CS term may allow for more freedom in the induced axion EM potential.} Thus, for our simple effective theory, we shall proceed with the Lagrangian (\ref{eq: EMLagr}).

This $\sim\vartheta F \tilde{F}$ term provides an additional contribution to standard Maxwell equations from the following ``Witten current" \cite{sikivie_interaction_1984, huang_structure_1985, dine_remarks_2023, wilczek_two_1987, kogan_axions_1993, lee_topological_1987}:
\begin{eqnarray}
\label{eq: jCS}
j_{CS}^{\mu} = \frac{\alpha}{2\pi}\partial_{\nu}\left(\vartheta\tilde{F}^{\mu\nu}\right).
\end{eqnarray}
Evidently, at constant $\vartheta$ this is a total derivative, however it now represents a conserved current $\partial_\mu j^\mu_{CS}=0$. Since we will discuss the effects of defects (e.g. magnetic monopoles) as a source of the EM fields in subsequent sections, we must also add the (classical) matter contributions to the Maxwell equations, which become
\begin{eqnarray}
&&\partial_{\nu}\left(F^{\mu\nu}-\frac{\alpha}{2\pi}\vartheta\tilde{F}^{\mu\nu}\right)
=j^{\mu}_E,\\
&&\partial_{\nu}\tilde{F}^{\mu\nu}=j^{\mu}_M,
\end{eqnarray}
where $j^{\mu}_E$ and $j^{\mu}_M$ are conserved electric and magnetic current respectively. To write this explicitly using $\vec{E}$ and $\vec{B}$,
\begin{eqnarray}
\label{eq: maxwell}
&&\vec{\nabla}\cdot\vec{E}+\vec{\nabla}\cdot\left(\frac{\alpha}{2\pi}\vartheta\vec{B}\right)=\rho_E,\\
&&\vec{\nabla}\times\vec{B}-\frac{\partial}{\partial t}\vec{E}
-\frac{\partial}{\partial t}\left(\frac{\alpha}{2\pi}\vartheta\vec{B}
\right)+\vec{\nabla}\times\left(\frac{\alpha}{2\pi}\vartheta\vec{E}\right)=\vec{j}_E,\\
&&\vec{\nabla}\times \vec{E}+\frac{\partial}{\partial t}\vec{B}=\vec{j}_M,\\
&&\vec{\nabla}\cdot\vec{B}=-\rho_M.
\end{eqnarray}


\section{The Monopole Bag}
\label{sec: mbag1}

Monopoles are magnetic charges that are ubiquitous in unified theories \cite{vilenkin_cosmic_2001}. Certainly, there is a range of models in which one may expect these objects to appear \cite{rajantie_magnetic_2003, kasuya_topological_1998}, however, as before we shall introduce a particular case for clarity. The 't Hooft-Polyakov monopole \cite{hooft_magnetic_1974, polyakov_particle_1974} is a topological soliton formed from symmetry breaking in Yang-Mills theories such as the breaking of some hidden $SU(2)_{\text{H}}$ gauge theory into (an eventual) $U(1)_{\text{EM}}$ \cite{baek_hidden_2014, khoze_dark_2014}. This is an example scenario where monopole formation might occur post-inflation, and thus be phenomenologically relevant. Unlike the Dirac monopole, there is no cosmic string attached. Thus, for a Higgs scalar $\Phi$ in the triplet representation of $SU(2)_{\text{H}}$ with symmetry breaking scale $\xi$, a simple 't Hooft-Polyakov monopole may be characterised by a typical Lagrangian \cite{preskill_magnetic_1984}
\begin{equation}
\label{eq: Lmon}
    \mathcal{L} = -\frac{1}{2}D_\mu\Phi^a D^\mu\Phi^a - \frac{\lambda}{4}\left(\Phi^a\Phi^a-\xi^2\right)^2 - \frac{1}{4}F^a_{\mu\nu}F^{a\mu\nu}.
\end{equation}
The core of the monopole is said to be a magnetically charged region of size $M_{m}^{-1}$ (with $M_m$ being monopole mass) respecting original symmetries prior to the breaking from which the monopole was formed \cite{preskill_magnetic_1984}. While we emphasise that our study will not concretely deal with nor necessitate an analysis of the core physics of monopoles, and the parameter space we will access in §4 may deviate from certain standard cases such as GUT, one could consider the above a fundamental setup motivating subsequent discussions.

\subsection{The Witten Effect}

It is known that particles with both magnetic ($q_m$) and electric ($q$) charge must satisfy the Dirac-Schwinger-Zwanziger (DSZ) quantisation condition \cite{dirac_quantised_1931, schwinger_magnetic_1966, zwanziger_quantum_1968}
\begin{equation}
\label{eq: dsz}
    q_1q_{m2} - q_2q_{m1} = 2\pi m.
\end{equation}
From this, we can see that the minimum charge required for a monopole to be quantum mechanically consistent in an EM theory is from taking $m=1$ and $q=e/2$ (minimum non-trivial electric charge of a 't Hooft-Polyakov monopole \cite{cheng_gauge_2011}), such that
\begin{equation}
\label{eq: gmin}
    q_m^{\text{min}} = \frac{4\pi g}{e}.
\end{equation}
Magnetic coupling will be taken to be $g=1$, as is fitting for our scenario in which $N_{\text{DW}}=1$ \cite{kogan_axions_1993}. It was then shown in the pioneering work by Witten \cite{witten_dyons_1979} that the presence of a variable, CP-violating $\vartheta$ constrains the permitted values for electric charge on a monopole. He showed that the charge on a monopole in an axionic background is
\begin{equation}
\label{eq: dyonch}
    q = ne + e\frac{\vartheta}{2\pi}g, \quad n\in\mathbb{Z}
\end{equation}
when taking $q_m=q_m^{\text{min}}$. Monopoles, by definition, possess no electric charge if $\vartheta=0$ (i.e. monopoles are the states with $n=0$), while dyons are objects with both electric and magnetic charge such that if $\vartheta=0$, dyons have integer charge $q=ne$. Witten's result therefore demonstrates that given non-zero $\vartheta$, monopoles possess dyonic excitations with electric charge $e\vartheta/2\pi$. Therefore, the $2\pi$ periodicity of $\vartheta$ allows shifting between different dyonic states, and renders certain states physically equivalent, e.g. $\{n=0,\vartheta\}$ and $\{n=1,\vartheta-2\pi\}$.

It is thus evident that a magnetic monopole traversing an axion domain wall retains the same magnetic charge but gains electric charge, which corresponds to both a mass transmutation and an apparent non-conserved change in the monopole's charge. The EM behaviour that allows one to rectify the latter lies in the overall conservation in the axion domain wall-EM field system, which is achieved by the realisation of an induced apparent charge on the domain wall sourced by the conserved Witten current (\ref{eq: jCS}). It follows that the induced charge and current are explicitly given by
\begin{gather}
        j^0 \sim \nabla\vartheta\cdot\vec{B}, \\
        \vec{j} \sim \dot \vartheta \vec{B} + \nabla \vartheta \times \vec{E}.
\end{gather}
A closed axion domain wall containing a monopole thus appears to be an effective dyonic state which, upon the removal of the enclosed monopole, returns to the overall charge conservation-obeying state of an uncharged axion domain wall and a dyon. This state is termed the ``monopole bag" \cite{kogan_kaluza-klein_1993, kogan_axions_1993}; let us first examine its known behaviour in flat space.

\subsection{The Monopole Bag in Flat Space}
\label{sec: mbag}

\begin{figure}
    \centering
    \includegraphics[width=0.8\linewidth]{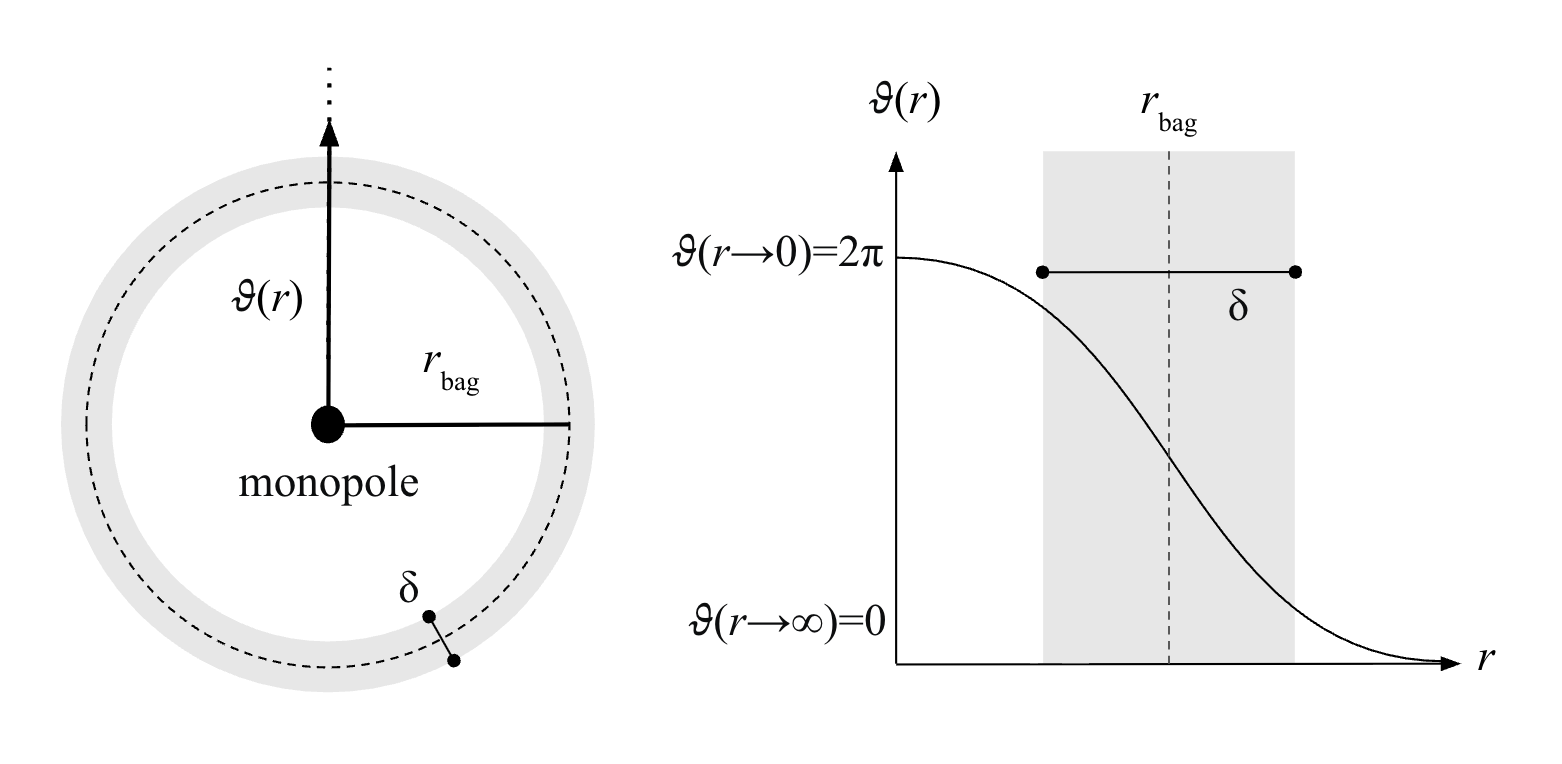}
    \vspace{-15pt}
    \caption{Visualisation of the monopole bag for choice of boundary values $\vartheta(r\rightarrow0)=2\pi,\text{ } \vartheta(r\rightarrow\infty)=0$, and a central monopole with size $r_{\text{core}}\ll r_{\text{bag}}$, $r_{\text{bag}}$ being the equilibrium radius of the external axion domain wall with thickness $\delta$.}
    \label{fig: bags}
\end{figure}

Now let us place a monopole in an axion domain wall as in Fig. \ref{fig: bags} and demonstrate that the monopole bag may form as a possible stable configuration. Suppose that there is an electrically neutral monopole enclosed within an axion domain wall. As the axion profile of which obeys $\vartheta(r\rightarrow\infty)=0$ and  $\vartheta(r\rightarrow0)=2\pi$, the central monopole is in fact a dyon with electric charge $q = e\vartheta/2\pi = e$ by taking $n=0$ in (\ref{eq: dyonch}). There are now radial electric and magnetic fields originating from the axion domain wall and the dyon (we have applied $g=1$ and $q_m=q_m^{\text{min}}$ for simplicity but without loss of generality in our conclusions):
\begin{equation}
\label{eq: Brnew}
\vec{B}(\vec{r})=\frac{q_m}{4\pi}\frac{\vec{r}}{r^3}\to B_r(r)=\frac{1}{er^2}
\end{equation}
\begin{equation}
\label{eq: ErK}
    \vec{E}(\vec{r})=-\frac{e^2}{8\pi^2}\vartheta\vec{B} + e\frac{\vec{r}}{r^3} \to E_r = \frac{e}{4\pi r^2} \left( 1 - \frac{\vartheta}{2\pi} \right)
\end{equation}
The system is thus an effective dyonic state with the equation of motion
\begin{equation}
\label{eq: thetaeom}
    0 = f_a^2\frac{1}{r^2}\frac{\partial}{\partial r}\left(r^2\frac{\partial\vartheta}{\partial r}\right) - m_a^2f_a^2\sin\vartheta + \frac{e^2}{32\pi^3r^4}\left(1-\frac{\vartheta}{2\pi}\right)
\end{equation}
where we have applied the condition of a time-independent, spherically symmetric field configuration. This matches an effective potential \cite{fischler_dyon-axion_1983}
 \begin{equation}
     V(\vartheta) = m_a^2f_a^2(1-\cos\vartheta) + \frac{e^2\vartheta^2}{32\pi^2}\left( 1-\frac{\vartheta}{2\pi} \right)^2.
 \end{equation}
Minimising this potential with respect to $\vartheta$, thus the equation of motion (\ref{eq: thetaeom}) can be solved to give
\begin{equation}
\label{eq: bag}
    \vartheta(r) = 2\pi \left(1 - e^{-r_{\text{bag}}/r}\right)
\end{equation}
with the stable bag radius being
\begin{equation}
    r_{\text{bag}} = \frac{e}{8\pi^2f_a}.
\end{equation}
The solution presented above can therefore be seen as a minimisation of total energy within the system, given that the right-hand side of (\ref{eq: thetaeom}) is the classical effective $\vartheta$ action. 

From the expression (\ref{eq: bag}), it is clear that the $\vartheta$ goes exponentially to $2\pi$ at small $r$ and to $0$ at large distances, outpacing the quadratic variation of the electric field. The electric field is therefore vanishing in the small $r$ limit, which corresponds to a complete screening of dyonic charge: outside of the domain wall, $E_r=e/(4\pi r^2)$, meaning the central dyonic charge contribution is entirely hidden for our physically motivated choice of $g=N_{\text{DW}}=1$. If one were to achieve screening for higher $g$, one would require $\vartheta=2\pi N/N_{\text{DW}}$ ($N\in\mathbb{Z}$) in place of our simple $2\pi$ periodicity in $\vartheta$. Notably, the electrostatic field energy $\vec{E}^2/2$ diverges for small $\vartheta$ values near the central dyon, therefore the dyon effectively ``repels" the domain wall, causing a strong localisation of $\vartheta=2\pi$ and sharp fall-off at $r_{\text{bag}}$. Note that changing the boundary conditions (e.g. taking instead $\vartheta(r\rightarrow\infty)=2\pi, \vartheta(r\rightarrow0)=0$) would not compromise the physical arguments reviewed in this section. To further contextualise the boundary conditions we have chosen, it has been noted (e.g., in \cite{kogan_axions_1993}) that typically, ordinary dyons should not themselves be classically stable due to (\ref{eq: thetaeom}) possessing an EM source term $\vartheta\vec{E}\cdot \vec{B}$ which drives $\vartheta$. However, in the present setup we see that the presence of an axion domain wall screens the interior dyonic charge completely and permits a spatially varying, static solution of the $\vartheta$ profile to be maintained, such that the final state can possess dyonic quantum numbers $q=g=1$ yet remain stable.

To assess the stability of the above choice of boundary condition, let us briefly discuss available decay channels. It is known that monopoles and dyons are capable of  pair producing light, charged states ($e^{\pm}$, $W^{\pm}$) via the Schwinger effect, as well as decaying into charged fermions and dyons \cite{gorsky_schwinger_2001, monin_monopole_2005}. We may focus on this perturbative, microscopic behaviour, though it arises from non-perturbative instanton processes within the monopole core gauge sector. In particular, $W^{\pm}$-production is understood to be possible once the electric field strength at the surface of a dyon (or, we extrapolate reasonably, an effective dyon in a monopole bag scenario) is sufficient for decay into a $W^{+}$ and a dyon with $q-1$. The dynamics of this process in an axion-monopole system with time-varying $\vartheta$ was studied in \cite{jeong_dissipation_2023}; while the work confirms the possibility of this dissipative process, it does not present a mechanism by which a topologically protected configuration of an axion domain wall and a monopole may unwind. The $W^\pm$ pair may be either standard model or in the central monopole gauge group e.g. $SU(2)_{\text{H}}$, but not one from each as we may generally assume there is no coupling between the two sectors. Under this framework, our stability statements may be supported. Further, light species emitted near the core region cannot perform the screening of dyon electric fields, which remains an essential property of the axion domain wall. As discussed in \cite{hook_dyonic_2025}, the bound state/core formation of light charged fermions has a rather flat profile, extended over a wide region with sharp damping at the radius $1/m$, and is unable to effectively screen the central dyonic charge. These circumstances may therefore support stability of the monopole bag state by virtue of the perfect screening supplied by the axion profile, and justify the Witten-inspired boundary condition that we can assume to be stable, even in the case where such a cloud is formed around the monopole bag. The constraint on the presence of these dynamical processes is often summarised as a bound on the dyon's size and mass, which in turn translates to an upper limit on the mass of the monopole bag \cite{kogan_kaluza-klein_1993}. Thus, for our purposes, some level of stability of the monopole bag structure against decay can be implied, though we must leave full analysis for future work. The particulars of permitted channels are an ongoing topic of research where conclusions are expectedly dependent on the parameters of each scenario; we do not further pursue this discussion beyond the above outline of the standard conditions.


\section{The Axion-Monopole System in Curved Spacetime}
\label{sec: amcurve}

We would now like to seek a gravitating final state that one may expect to be solitonic, or to resemble a dyonic black hole solution with a possible regular centre in our simple construction using matched spacetimes. In standard, existing studies of regular black holes \cite{hayward_formation_2006, mbonye_nonsingular_2005, davies_nonsingular_2025, kleihaus_stationary_2008}, a non-singular centre is often created through either the gravitational theory employed preventing the formation of a singularity even in the event of catastrophic collapse, or exotic matter properties that somehow counteract the collapse at the centre. Let us turn our attention to the latter case. Should one imagine an object (such as that proposed by Hayward \cite{hayward_formation_2006}) for which spacetime is asymptotically flat at large $r$ and de Sitter (dS) as $r\rightarrow0$, the effective cosmological constant in the centre may oppose singular collapse. Intriguingly, the monopole has also been proposed as a self-gravitating, stable\footnote{See e.g. \cite{kain_are_2019} for further discussion on stability.} object possessing dS geometry within its core and an asymptotically flat metric in its exterior \cite{arreaga_stability_2000, pazameta_general_2012, ayon-beato_bardeen_2000}. We therefore begin to ask the question of whether it would be possible to construct a stable compact object starting from a monopole enclosed in a naturally collapsing axion domain wall by analysing the validity of the final state.

There are certain caveats to the stability of dyonic regular black holes that deserve to be briefly addressed prior to delving into the potential of our upcoming analysis serving as an initial indication of a novel construction. It has recently been found in \cite{felice_instability_2025} that a Laplacian instability is expected to develop, however, as shown in \cite{junior_dyonic_2025, felice_exotic_2025}, there are also additional considerations that would lend longevity to such an object, namely through scalar potentials and non-trivial scalar-vector EM couplings respectively. In particular, we note here that the effect of an axionic coupling with a well-motivated scalar potential is yet to be considered in current literature to our best knowledge. Another interest in our ensuing study is therefore to build a simple model that may be refined into a dyonic black hole-type object without the need for invoking either exotic gravity theories nor higher order EM deviations from familiar and established theories.

\subsection{Geometrical Setup}

Let us write down in full the low-energy Lagrangian of our theory in terms of the axion field $a(r)$ with gravity minimally coupled by an Einstein-Hilbert term with Ricci scalar $\mathcal{R}$:
\begin{equation}
\label{eq: Lsimp}
    \mathcal{L} = \frac{M_{Pl}^2}{2}\mathcal{R} - \frac{1}{2}\partial_\mu a(r) \partial^\mu a(r) - m_a^2f_a^2\left[ 1-\cos\left( \frac{a(r)}{f_a} \right) \right] - \frac{1}{4}F_{\mu\nu}F^{\mu\nu} + \frac{\alpha a(r)}{8\pi f_a}F_{\mu\nu}\tilde{F}^{\mu\nu}.
\end{equation}
As we will be exclusively interested in closed axion domain walls, we will henceforth deal with spherically symmetric axion configurations only, with all future explicit $r$-dependence dropped for notational simplicity. The EM field tensor is defined using the vector field $A_\mu$ as $F_{\mu\nu} = \partial_\mu A_\nu - \partial_\nu A_\mu$, where we may write the one-form \cite{ramirez-valdez_dyonic_2023}
\begin{equation}
\label{eq: A1form}
    A_\mu dx^\mu = A_0(r)dt - q_m\cos\theta d\phi.
\end{equation}
The $t$-component $A_0$ is proportional to the electric charge of the system $q$, while the $\phi$-component encodes magnetic charges. All EM fields will also be assumed spherically symmetric from now on. By the topological nature of the $\sim(a/f_a)F\tilde{F}$ term, it possesses no direct coupling with gravity nor contribution to the energy-momentum tensor, which is expressed as
\begin{equation}
    T^{\mu}{}_{\nu} = \left[ - \frac{1}{2} \partial^{\beta} a \partial_{\beta} a - m_a^2 f_a^2 \left( 1 - \cos\left( \frac{a}{f_a} \right) \right)  - \frac{1}{4} F_{\beta\kappa} F^{\beta\kappa} \right] \delta^{\mu}_{\nu} + \partial^{\mu} a \partial_{\nu} a - F^{\mu\beta} F_{\beta\nu}.
\end{equation}
Thus, the gravitational effects from $(a/f_a)F\tilde{F}$ may only appear in the field equations; the equations of motion for our given Lagrangian are then
\begin{equation}
\label{eq: Leom}
    \Box a - m_a^2f_a\sin\left(\frac{a}{f_a}\right) + \frac{\alpha}{8\pi f_a}F_{\mu\nu}\tilde{F}^{\mu\nu}=0,
\end{equation}
\begin{equation}
\label{eq: EMeom}
    \nabla_\nu\left( F^{\mu\nu} - \frac{\alpha a}{2\pi f_a}\tilde{F}^{\mu\nu} \right) = 0.
\end{equation}
Note that while the right-hand side of the equations of motion for the EM tensors are $0$ only outside the monopole core, given we have opted to simplify our problem by not treating the core interior explicitly, the above equations of motion will suffice for the external regions analysed in subsequent sections. 

For the geometry exterior to the monopole core, it is well-known that one must employ the following metric ansatz to describe an asymptotically flat spacetime inhabited by non-trivial field configurations:
\begin{equation}
\label{eq: Nansatz}
    ds^2 = -C(r)N(r)dt^2 + C^{-1}(r)dr^2 + r^2d\Omega^2.
\end{equation}
The yet undefined functions $N, C$ are constrained such that the ansatz corresponds to a Reissner-Nordstr\"{o}m (RN) spacetime at asymptotic distances due to our charged construction, and all functions should depend only on the radial coordinate $r$, as is consistent with our assumed spherical symmetry. This setup for our axion model with gravity shares essential points with the pioneering work by Lee-Weinberg on charged black holes with scalar hair \cite{lee_charged_1991}. In addition to their well-known case, we have included the axion potential and particle physics-motivated boundary conditions; under our original motivation, we analyse both the soliton-like and black hole-like static states.

A description of the core region of the monopole can be found in various literature \cite{preskill_magnetic_1984,tong_tasi_nodate,weinberg_classical_1992, bolognesi_magnetic_2011}. For our purposes, it suffices to summarise that given a typical Mexican hat potential for monopole-conducive symmetry breaking as described in (\ref{eq: Lmon}), the interior of the monopole may be described by dS-like geometry:
\begin{equation}
\label{eq: dSansatz}
    \begin{gathered}
        ds^2=-(1-H^2r^2)dt^2+(1-H^2r^2)^{-1}dr^2+r^2d\Omega^2.
    \end{gathered}
\end{equation}
$H$ is the Hubble parameter $H^2=8\pi/3\cdot V(0)=2\pi\lambda\xi^4/3$, and the above model is accurate for e.g. an $\text{SU}(2)$ gauge symmetry broken to EM $\text{U}(1)$. Within the core, the original symmetries are respected, such that in the $\text{SU}(2)\rightarrow\text{U}(1)$ example, we can expect Abelian, Maxwell-like gauge fields outside the core. The values of mass and core radius are of course strongly model-dependent, however we can make the general statement that should the core radius be much larger than the monopole's Compton wavelength, we may treat the object as classical. More relevant to our purpose is the treatment of the geometries in this problem. The core will be taken to be a spherical shell of vanishing thickness, behaving as a timelike hypersurface across which Darmois-Israel junction conditions are applied to permit matching between (\ref{eq: Nansatz}) and (\ref{eq: dSansatz}). Namely, for an induced metric $h_{\mu\nu}$ on the shell, an extrinsic curvature $K_{\mu\nu}$, and surface energy-momentum $S_{\mu\nu}$ \cite{tanahashi_spherical_2015}:
\begin{equation}
\label{eq: DIcond}
    \begin{gathered}
        \left[h_{\mu\nu}\right]=h^+_{\mu\nu}-h^-_{\mu\nu}=0,\\
        \left[K_{\mu\nu}\right]=\frac{1}{M_{Pl}^2}\left( -S_{\mu\nu}+\frac{1}{2}h_{\mu\nu}S \right).
    \end{gathered}
\end{equation}
We emphasise that these assumptions are simple and thus not sufficient for a more complete UV assessment; where black hole geometries occur in upcoming sections, this above scenario must be re-examined with care if a rigorous proof were to be sought.

Thus, one might expect that the metric (\ref{eq: Nansatz}) behaves as
\begin{equation}
\label{eq: ideal}
    \begin{gathered}
        C(r\rightarrow\infty) = 1 - \frac{2M}{r} + \frac{Q^2}{r^2} + \mathcal{O}(r^{-3}),\\
        C(r\rightarrow0) = 1 - H^2r^2 + \mathcal{O}(r^{4}),\\
        N(r\rightarrow\infty) = 1,\\
        N(r\rightarrow0) = N_0,
    \end{gathered}
\end{equation}
which corresponds to RN charged by $Q$ asymptotically and dS with $H^2 = (8\pi/3)V(r=0)$ in the centre. The boundary conditions on $N$ are such that at $r_{\text{core}}$, one should be able to match an external asymptotically flat spacetime with some constant $N_0$ to the internal dS spacetime. Solving the Einstein equations
\begin{equation}
\label{eq: EE}
    G^\mu_\nu = \frac{8\pi}{M_{Pl}^2}T^\mu_\nu,
\end{equation}
it is easy to verify that the $(\theta, \theta)$ and $(\phi,\phi)$ components expectedly coincide. We may treat the ensuing stress-energy as that of an anisotropic effective fluid. Factors of $M_{Pl}$ will be implicit hereafter for cleaner equations, i.e. $G=1$ will now be taken notationally only.

In order to begin solving the system, we can first constrain the form of $A_0$ by its equation of motion
\begin{equation}
\label{eq: A0N}
    A_0'(r) = \frac{\sqrt{N(r)}}{r^2}\left(q-\frac{\alpha q_m a(r)}{2\pi f_a}\right),
\end{equation}
and derive the constraint on $a$ and $N$
\begin{equation}
\label{eq: a'C}
    a'^2=\frac{N'}{8\pi rN}
\end{equation}
by taking the difference between the $(t,t)$ and $(r,r)$ components of the Einstein equations. This result demonstrates the tie between the axion profile and the gravitational redshift $N$, where the non-trivial form of one necessarily predicts a variation in the other. Now writing down the equation of motion for $C$ and the (t,t) component of the Einstein equations respectively:
\begin{equation}
\label{eq: aCeom}
    \begin{gathered}
        -4\pi^{2} f_a^{3} m_a^{2} r^{4} \sin\left(\frac{a}{f_a}\right) + 4\pi^{2} f_a^{2} r^{4} C a'' + 16\pi^{3} f_a^{2} r^{5} a'^{3} C \\ + 4\pi^{2} f_a^{2} r^{3} \left(r C' + 2C\right) a' - \alpha^{2} q_{m}^{2} a + 2\alpha q q_{m} \pi f_a = 0,
    \end{gathered}
\end{equation}
\begin{equation}
\label{eq: aCtt}
    -\frac{1}{2}a'^2C -m_a^2f_a^2\left[1-\cos\left(\frac{a}{f_a}\right)\right] -\frac{q_m^2}{2r^4}-\frac{A_0'^2}{2N} = \frac{r C' + C - 1}{8\pi r^{2}},
\end{equation}
by substitution with (\ref{eq: A0N}) and suitable simplification, these now serve to constrain the relation between $a$ and $C$. Equations (\ref{eq: A0N}-\ref{eq: aCtt}) are highly non-linear, therefore we must seek a numerical solution. To do so, we must also impose the boundary conditions $a(r_{\text{core}})=2\pi f_a$, $a(r\gg r_{\text{bag}})=0$ as in §\ref{sec: adw}, and on metric components, the same requirements as (\ref{eq: ideal}) for asymptotic flatness ($C(r\gg r_{\text{bag}})=1$) and regularity ($C(r_{\text{core}})=1$) apply. Numerical methods suffer certain drawbacks in situations with strongly divergent terms, such as the high powers of $r$ in our equations. We will therefore impose a cut-off at small $r$; this is a general choice we are free to make, as we are continuing to treat the monopole as an inert centre at $r \leq r_{\text{core}} < r_{\text{DW}}$. In addition to searching for metric function and axion field profiles in a final state, we will also plot the mass and the energy density,\footnote{While we refer to the quantity $|-T^t_t|$ as the ``energy density", where we deal with regions internal to the horizon in subsequent analysis, this quantity no longer corresponds to energy density in the physical sense (e.g. Fig. \ref{fig: strong}). To make it so would demand a full solution in a Kruskalised coordinate system, which would be extraneous to the behaviours---specifically $a$ profile shape---we wish to investigate and is thus omitted.}
\begin{equation}
\label{eq: rho}
    -T^t_t \equiv \rho_\text{tot} = \frac{1}{2}a'^2C + m_a^2f_a^2\left[1-\cos\left(\frac{a}{f_a}\right)\right] + \frac{(2\pi f_aq - \alpha aq_m)^2+4\pi^2f_a^2q_m^2}{8\pi^2f_a^2r^4},
\end{equation}
with the former being obtained from integration of the latter. In subsequent analysis, the axionic terms in (\ref{eq: rho}) will be referred to as $\rho_{\text{fields}}$ and used to isolate the domain wall contributions to total energy density $\rho_\text{tot}$, which is dominated by large EM terms at small scales.


\subsection{The Gravitating Monopole Bag}
\label{sec: sols}

\begin{figure}[!ht]
    \centering
    \includegraphics[width=\linewidth]{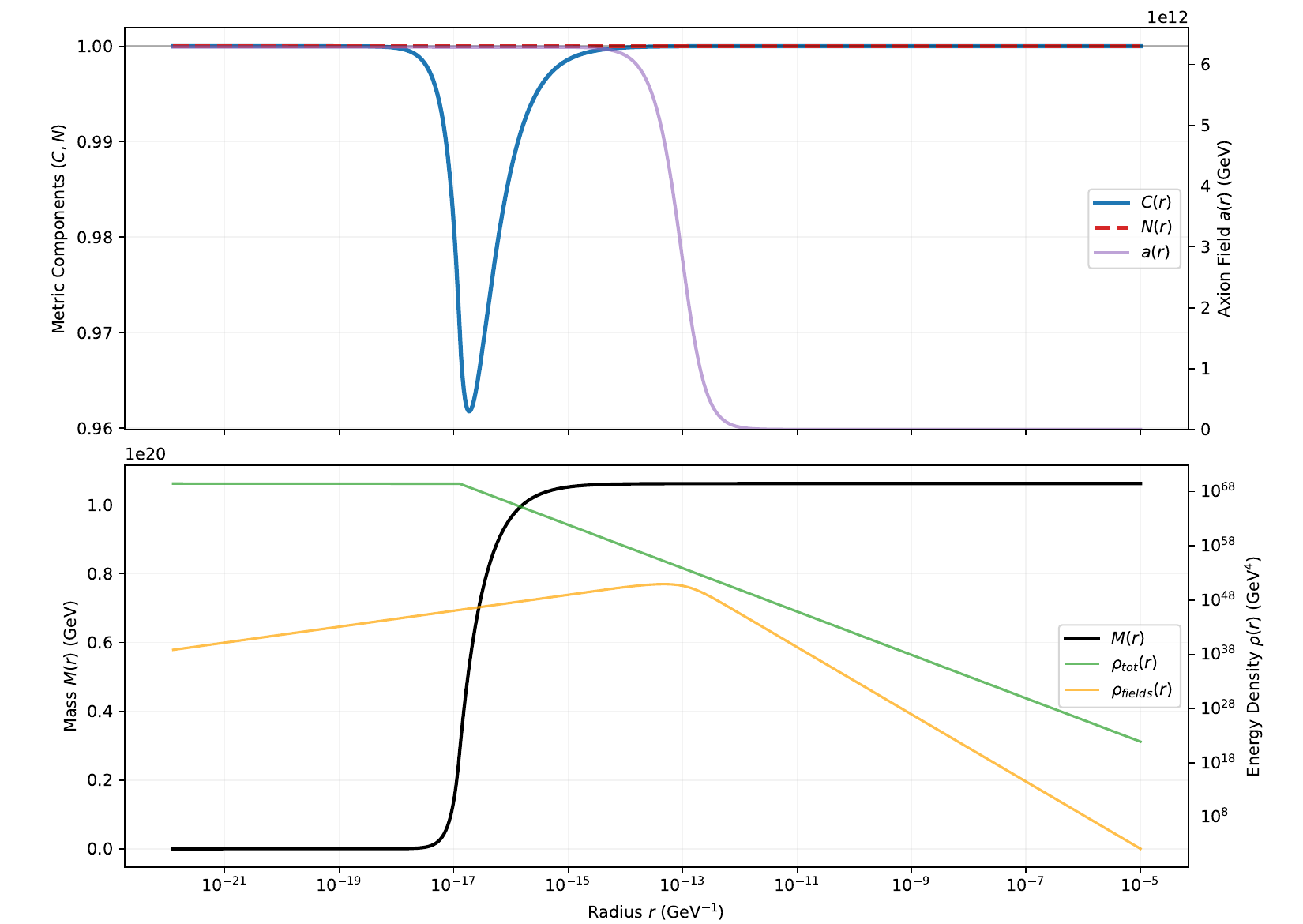}
    \caption{Plotting $C$, $N$, $a$, $\rho_{\text{tot}}$, $\rho_{\text{fields}}$, and $M$ given QCD axion scale parameters $f_a=10^{12}GeV$, $m_a=10^{-6}eV$, $q=1$, $q_m=4\pi$, $\alpha=1/4\pi$, $M_m=10^{18}GeV$. The monopole bag is entirely solitonic at these values, as shown by the horizonless, nearly flat metric functions.}
    \label{fig: reszout}
\end{figure}

We begin by analysing Fig. \ref{fig: reszout}, which shows the results for $f_a=10^{12}GeV$ and $m_a=10^{-6}eV$, corresponding to values typical of QCD axion models. Here, we imposed a monopole mass of $M_m=10^{18}GeV$ behaving as a constant localised to the core region $r_{\text{core}}\sim 1/(\alpha M_m) \ll r_{\text{DW}}$ as a choice that does not violate any present bounds \cite{rajantie_magnetic_2003, gould_magnetic_2017}. At these values, the main dip in the shape of $C$ reflects the solitonic property of the central monopole, while the axion domain wall does not give rise to any discernible gravitational contribution that can be seen to significantly impact the geometry. The total mass $M$ is therefore dominated by monopole mass and EM terms in (\ref{eq: rho}).\footnote{Note that the plateau of $M$ in Fig. \ref{fig: reszout} is $10^{20}GeV>M_m=10^{18}GeV$. This comes from our choice of $q_m=4\pi$, resulting in the term $q_m^2/2r^4$ in (\ref{eq: rho}) giving rise to this mass in the first decade from $r_{\text{core}}$. The choice of $q_m$ is a matter of convention in some sense; our choice was made to align with that in standard literature pertaining to similar setups involving 't Hooft-Polyakov monopoles, e.g. \cite{bai_hairy_2021}.} While $N$ seems to remain entirely flat at $N(r)=1$ for these values, this is not due to a flat $a$ profile as one may na\"{i}vely conclude from (\ref{eq: a'C}), but is instead a consequence of the relative scale between gravitational effects and the energy density of the wall. Equation (\ref{eq: a'C}) can be solved for $N$ as
\begin{equation}
\label{eq: Nsol}
    N(R)=e^{M_{Pl}^{-2}\int^\infty_R dr ra'^2},
\end{equation}
where we now explicitly reinsert $M_{Pl}$. It is thus clear that while the energy of the wall $\sim a'^2 \ll M_{Pl}^2$, $N$ remains very close to 1.

\begin{figure}[!ht]
    \centering
    \includegraphics[width=\linewidth]{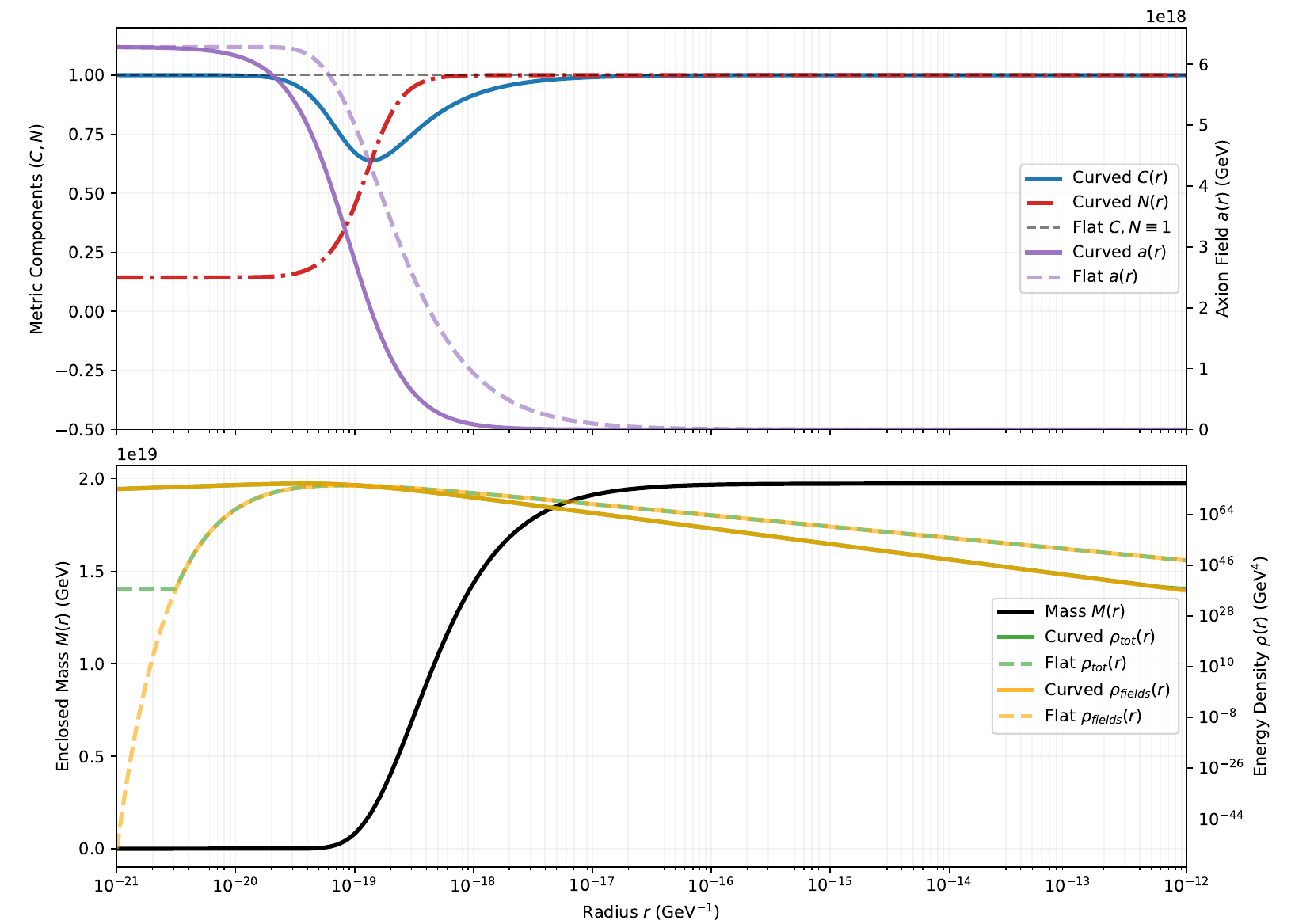}
    \caption{A plot for the strong field case where $f_a=10^{18}GeV$, $m_a=10^{-11}eV$, $q=1$, $q_m=4\pi$, $\alpha=1/4\pi$, $M_m=10^{10}GeV$. Curved (solid) and flat (dashed) space results are overlaid. Domain wall mass is now dominant over monopole mass, thus solely contributes to the geometry, causing a dip in the metric function $C$ and the expected redshift behaviour through a non-trivial $N$.}
    \label{fig: smallMm}
\end{figure}

We may isolate the pure domain wall effects by taking $f_a=10^{18}GeV$ and $m_a=10^{-11}eV$, then making monopole mass comparatively negligible by selecting $M_m=10^{10}GeV$ (Fig. \ref{fig: smallMm}). These axion parameters fall within the realm of more general ALP models, and correspond to a much more massive domain wall. This allows us to observe the behaviour of the domain wall more clearly, where previously its mass was subdominant to the monopole and its associated EM mass-energy. Indeed, the domain wall now impacts the shape of $C$ and $N$, the latter being particularly demonstrative of this as it attains a non-trivial profile in line with boundary conditions expected (\ref{eq: ideal}). The energy densities $\rho_{\text{fields}}$ and $\rho_{\text{tot}}$ coincide in each of flat and curved space scenarios plotted, where we observe a constant ``background" value at small $r$ in flat space $\rho_{\text{tot}}$, which can be identified as EM contributions. The total mass for curved space is plotted to demonstrate the recovery of a finite asymptotic value.\footnote{The scaling on the energy density profiles for flat and curved cases has been chosen such that the two are aligned and display in a useful way. In reality, if one were to impose some total mass equivalence between the two cases, there would be a distinct offset in $\rho_{\text{tot}}$ profiles to account for the flat space domain wall being thinner. However, we also point out that if one were to deal in reality, accretion/emission and the regulatory effect the gravitating scenario may generate in order to maintain balance between EM and gravity would also require a detailed analysis in order to determine true mass relations.} We notice that the wall in the curved spacetime appears both translated and distorted, as is clear from the differing energy density profiles in flat and curved space. Before delving further into observed behaviours, we note that the entire structure observed in Fig. \ref{fig: smallMm} lies within the monopole core region should we adhere to our estimate of $r_{\text{core}}\sim1/(\alpha M_m)$. However, the Witten effect is applicable irrespective of whether the theory is Abelian (which, one may recall, is only true strictly outside of the monopole core region), and taking the monopole core to resemble a thin domain wall i.e. a sharp boundary (separating regions with e.g. $SU(2)$ and $U(1)$ symmetries in the case of a 't Hooft-Polyakov monopole) is a particularly simplistic representation. Certainly, the Compton wavelength condition introduced is also only broadly applicable; in order to correctly derive the true size of the core region and accurately describe the physics within, one might require a UV model or at least a selection of a particular monopole configuration. Thus, given that equations not too dissimilar in structure to ours can still be reasonably applied, and we continue to impose a small $r$ to evade numerical infinities and regions projected to have the strongest UV effects, Fig. \ref{fig: smallMm} serves as an acceptable illustrative case for observing isolated domain wall behaviour in a gravitating monopole bag.

With the understanding of expected domain wall contribution gained above in mind, we will henceforth consider values satisfying $r_{\text{core}}<r_{\text{DW}}$ in order to treat the region of interest external to the monopole core with an effective theory in a more confidently physical manner. Consequently, we will begin using monopole masses far beyond that which can be formed via e.g. standard GUT symmetry breaking. However, we note that the ingredients required for the Witten effect to supply our monopole bag setup (namely, the non-trivial CS term and charge quantisation) do not demand a monopole Lagrangian (\ref{eq: Lmon}), a description of the core, or a GUT origin. While for our purposes we do not claim physicality, we highlight that there exist alternative possible sources of monopoles with a wider range of masses, such as those arising from higher dimensional theories where monopole mass is determined by geometry in place of Higgs VEV \cite{gross_magnetic_1983, sorkin_kaluza-klein_1983, diaconescu_d-branes_1997, verlinde_metastable_2006}. We also hereby draw attention to the highly relevant gravitating monopole construction by Breitenlohner \textit{et al}. \cite{breitenlohner_gravitating_1992}, a precedent case considering the possible consequences of heavy, gravitating monopoles further motivating upcoming discussion.

\begin{figure}[!ht]
    \centering
    \includegraphics[width=\linewidth]{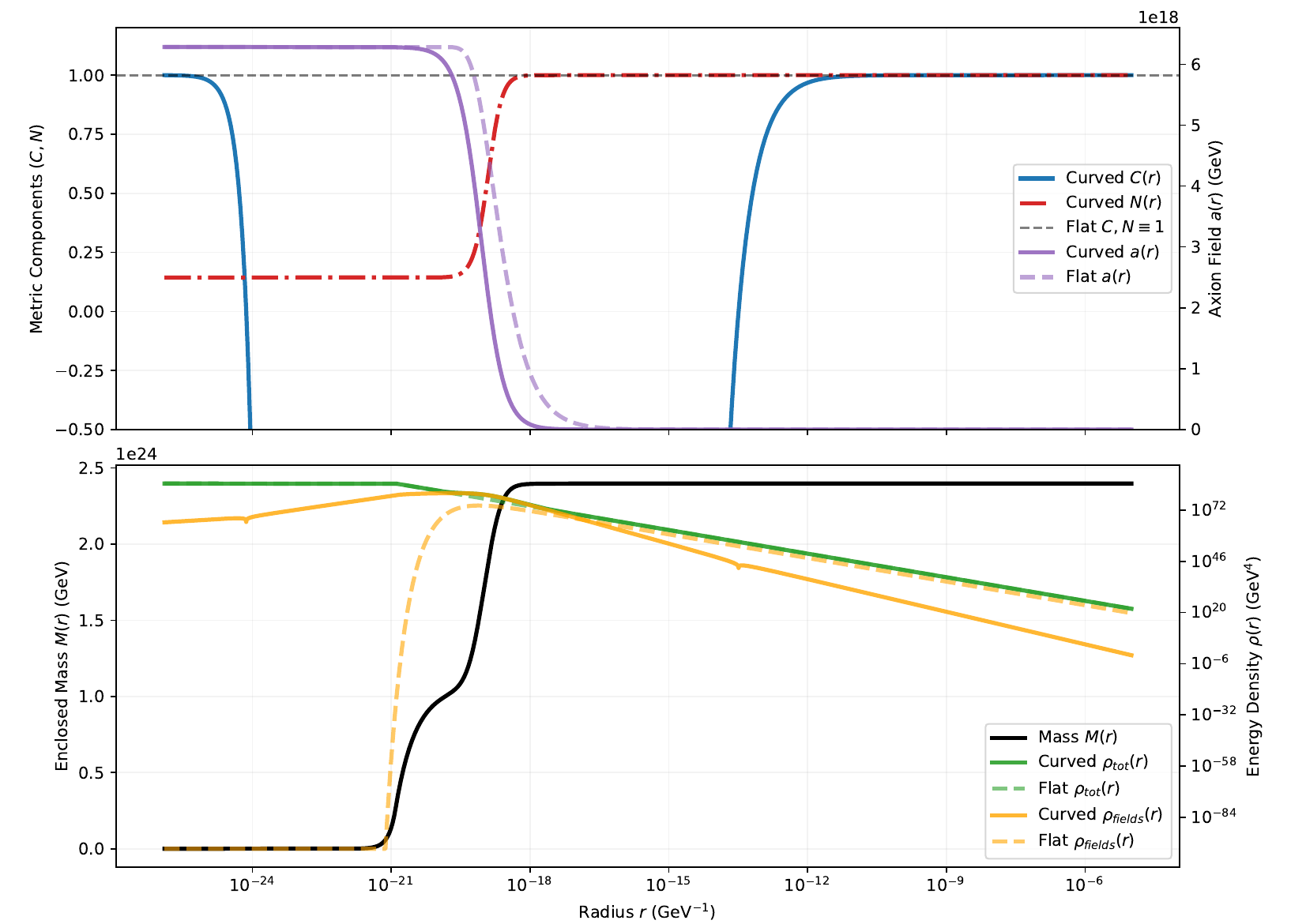}
    \caption{A plot for the strong field case where $f_a=10^{18}GeV$, $m_a=10^{-11}eV$, $q=1$, $q_m=4\pi$, $\alpha=1/4\pi$, $M_m=10^{22}GeV$. The domain wall mass is significant with respect to the total mass-energy of the system at these ALP parameters, and the monopole radius is now decreased such that the region of interest external to the core can be more accurately described by our classical, effective formalism.}
    \label{fig: strong}
\end{figure}

\begin{figure}[!ht]
    \centering
    \includegraphics[width=0.8\linewidth]{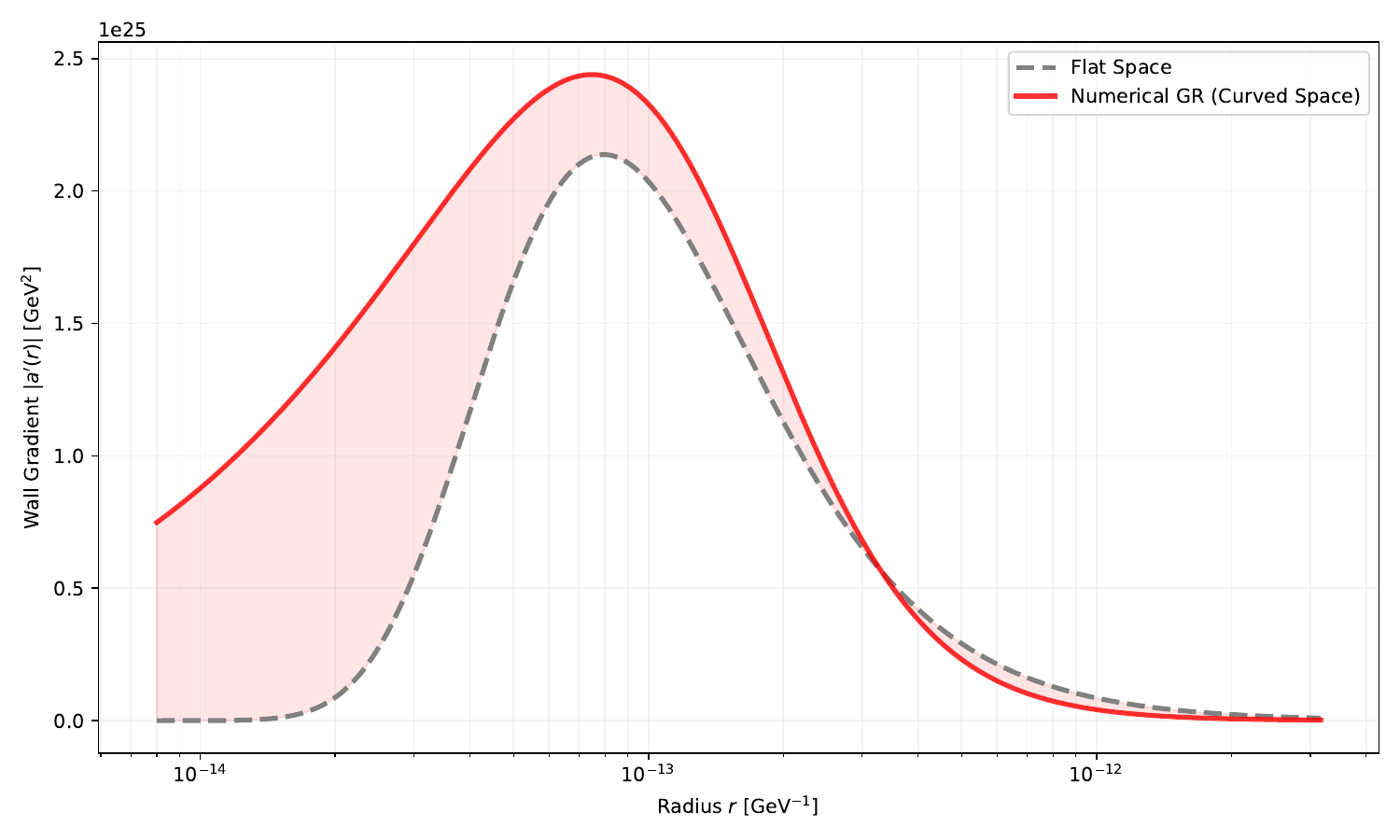}
    \caption{For $f_a=10^{18}GeV$, $m_a=10^{-11}eV$, $q=1$, $q_m=4\pi$, $\alpha=1/4\pi$, $M_m=10^{22}GeV$, the flat and curved space axion profiles ($a$ gradients) are compared. The axion profile in the curved case evades flatness at small $r$ but plateaus more sharply at large $r$ due to gravitationally-inflicted compression of the shape dominating at large radii, while EM effects prevent $a$ from rolling steeply away from $2\pi$ at small radii.}
    \label{fig: grad}
\end{figure}

Let us therefore consider a scenario with $f_a=10^{18}GeV$, $m_a=10^{-11}eV$, and $M_m=10^{22}GeV$ (Fig. \ref{fig: strong}). Firstly, $\rho_{\text{tot}}$ now receives a visible contribution from axion terms, and the total mass has become sufficient to create an event horizon. The active contribution of the domain wall to the geometry is now clear, as its mass has resulted in the event horizon being at much larger $r$ than would be otherwise expected, such that it now swallows most of the domain wall. From the plot of $\rho_{\text{fields}}$, we conclude that there is a non-zero energy density outside the horizon originating from the tail of the axion profile; since $C\simeq1$ outside the horizon, from the form of $\rho_{\text{tot}}$ given in (\ref{eq: rho}) we infer that there the gradient of $a$ must be non-zero, meaning the asymptotic value $a(\infty)=0$ has not been reached yet. By analogous reasoning, it is evident that at small $r$, the curved case axion profile is non-trivial until closer to the core than in the flat case. That is to say, the required limit $a(r\rightarrow0)=2\pi$ is reached more gradually in the curved case. This is evidence of the pronounced gravitational effects on the domain wall: in a gravitating system, the domain wall energy density seems to be smeared out in small $r$ regions. This effect is isolated in Fig. \ref{fig: grad}, in which we explicitly see that the gradient of the axion profile is steeper at small $r$ if gravity is active.

\begin{figure}[!ht]
    \centering
    \includegraphics[width=\linewidth]{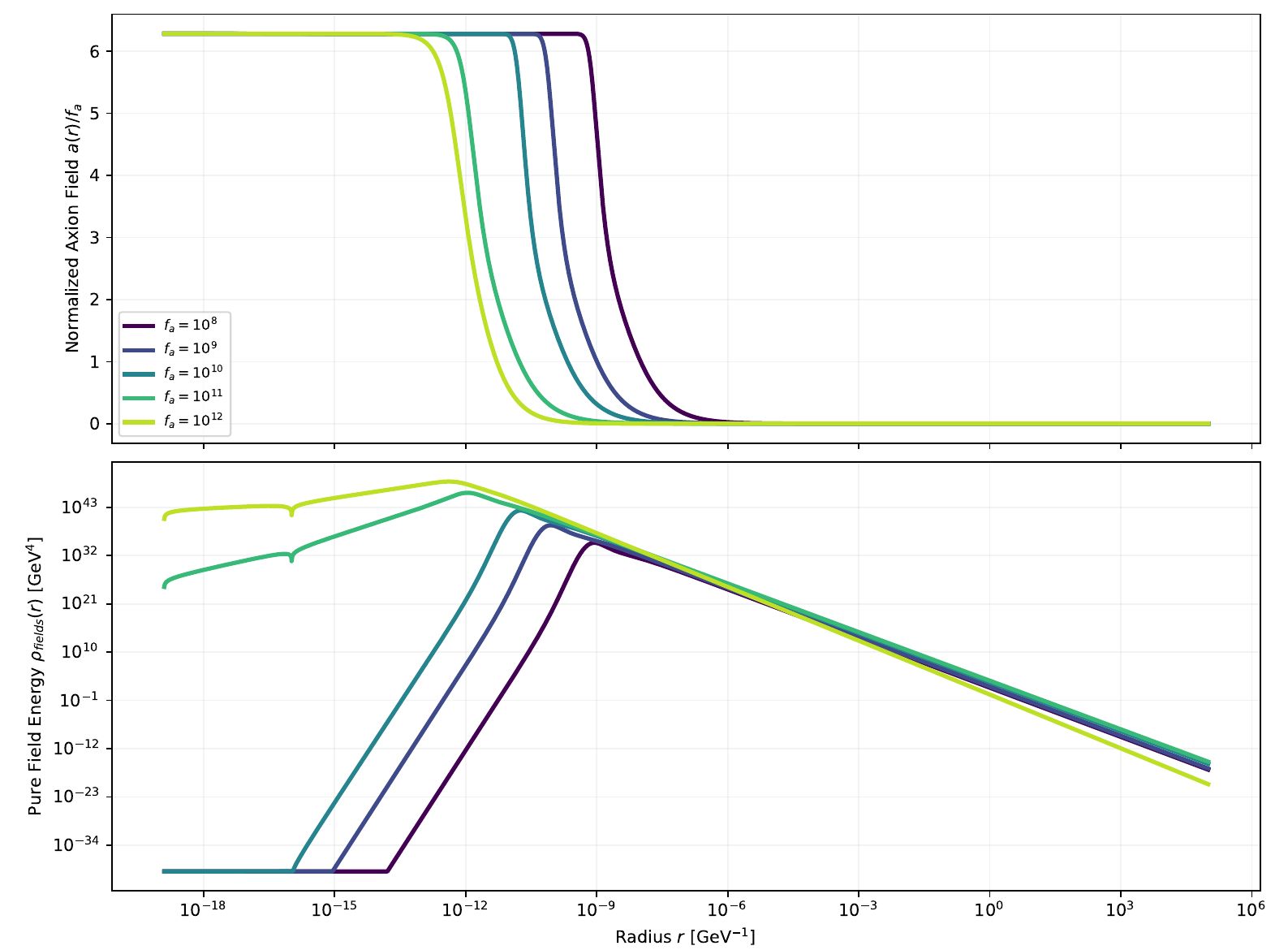}
    \caption{A range of $f_a$ was checked while maintaining constant values of $q=1$, $q_m=4\pi$, $\alpha=1/4\pi$, $M_m=10^{22}GeV$. At high $f_a$ i.e. large domain wall mass, gravitational effects are strong, which is reflected in large $\rho_{\text{fields}}$ (thus non-zero $a'$) being maintained at small radii indicating the wall is being compressed as far as EM effects permit.}
    \label{fig: fchange}
\end{figure}

Let us now vary the parameters of the problem within a larger range to further pinpoint the root of this deformation in the profile. In Fig. \ref{fig: fchange}, we vary only the value of $f_a$ (and $m_a$ correspondingly for particularly conservative estimates \cite{bernal_alp_2021}, though there is arguably no longer a particularly strict relation between the two for ALP models \cite{navas_review_2024}) to find that as the wall becomes more massive with larger $f_a$, the gravitational force working to crush the wall to a smaller radius is more prominent, such that $a'$ becomes non-zero (i.e. $a$ is forced to roll away from its inner boundary value of $2\pi$ towards its asymptotic value of $0$ earlier, as the domain wall is pressed inwards) at much smaller $r$. The extent to which this is allowed is limited by EM terms: the presence of the axion-monopole interaction results in a divergent cost in energy should $a(r\rightarrow r_{\text{core}})$ deviate significantly from $2\pi f_a$, as can be deduced from (\ref{eq: ErK}). Consequently, at small $r$, while strong gravitational forces drive $a$ inwards (successfully pushing $r_{\text{bag}}$ inwards, as seen by the peak location of $\rho_{\text{fields}}$ being decreased with increased $f_a$), EM terms force $a(r)\simeq 2\pi f_a$ to be maintained as much as possible until they have fallen sufficiently as $1/r^4$ to permit $a$ to drop towards $0$. The small cusps correspond to the location of the outer event horizon, which moves outwards as the total mass-energy of the system is increased at constant charge. The large $r$ tail is due to both gravitational and EM effects fading and the wall beginning to obey the well-known flat space sine-Gordon solution
\begin{equation}
    \vartheta\sim\arctan(e^{-m_ar})
\end{equation}
at sufficiently large radii. The fall-off is sharper in the strongly gravitating case due to the residues of gravity at larger $r$ finally dominating over EM forces, thus successfully compressing the wall.

\begin{figure}[!ht]
    \centering
    \includegraphics[width=\linewidth]{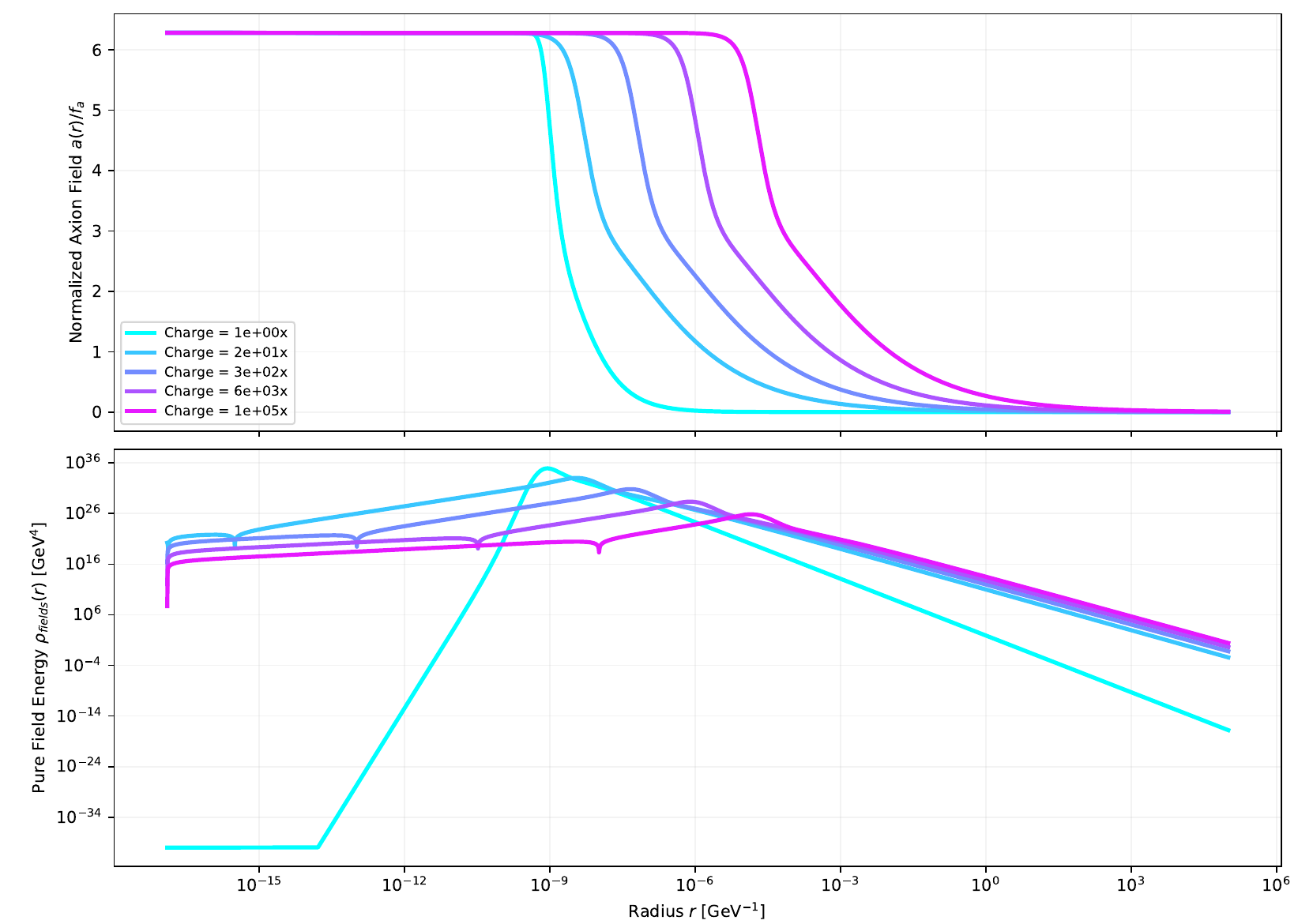}
    \caption{The charges $q$ and $q_m$ are increased by applying a multiplier on the baseline values taken so far while maintaining constant $f_a=10^{8}GeV$, $m_a=10^{-6}eV$, $M_m=10^{18}GeV$. As charge increases, the wall profile is seen to smear and move outwards due to strong outward EM pressure.}
    \label{fig: qchange}
\end{figure}

We can in fact show the opposite variation by changing the charges $(q,q_m)$, and thus the strength of the EM term. In Fig. \ref{fig: qchange}, as charge is increased, $r_{\text{bag}}$ moves outwards due to a larger EM term corresponding to a larger cost in energy for $a\neq2\pi f_a$ at small $r$. We also observe that the peaks in $\rho_{\text{fields}}$ shift down and right with increasing charge as the overall profile is flattened; with gravity becoming overpowered by EM terms, $a'$ is suppressed for longer, and the fall back to $a=0$ is correspondingly slow as reflected by higher-charge $\rho_{\text{fields}}$ profiles possessing a shallower tail at large $r$. However, the extended profile at smaller $r$ is also present at higher charge, as the higher charge EM terms boost the total mass-energy.

\begin{figure}[!ht]
    \centering
    \includegraphics[width=0.8\linewidth]{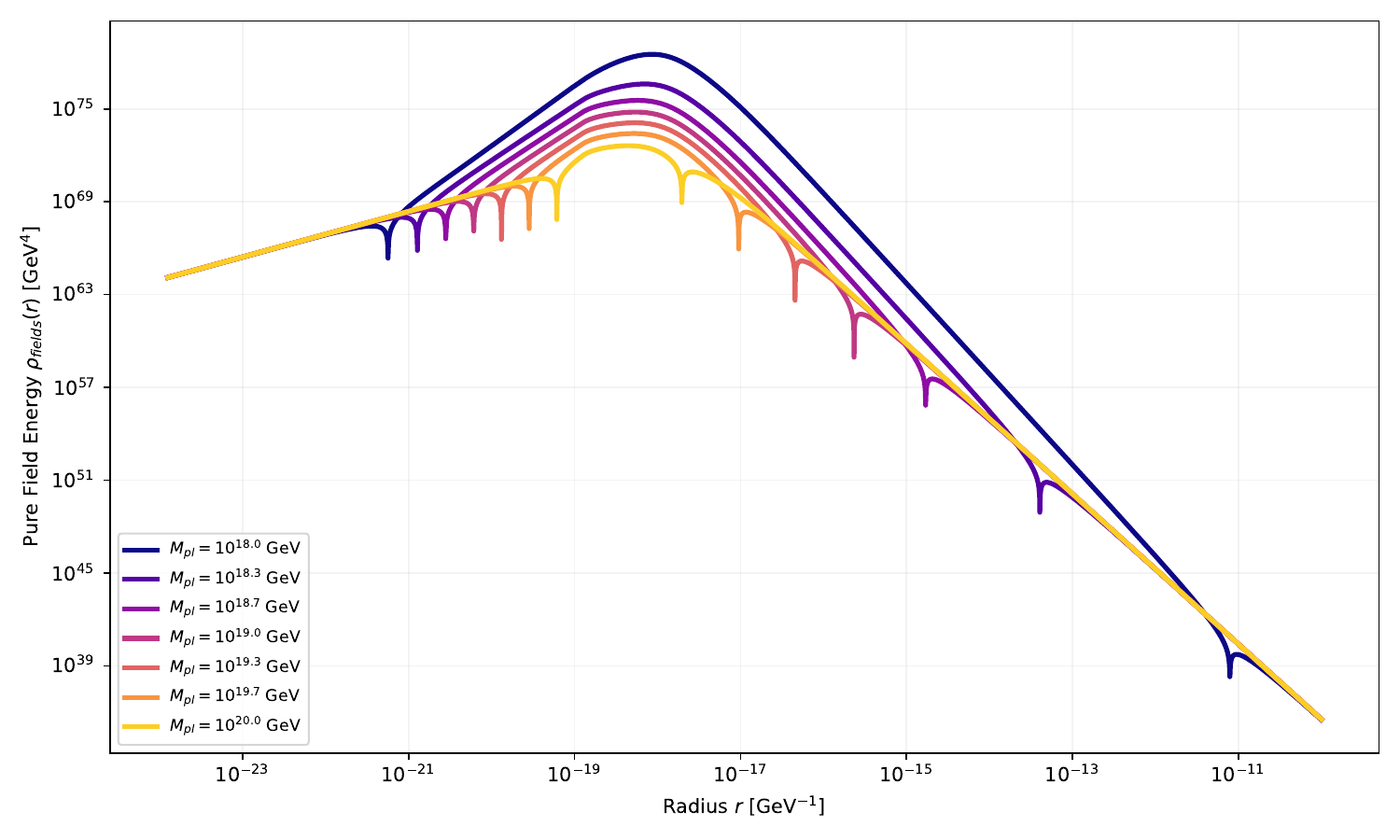}
    \caption{By changing $M_{Pl}$ (i.e. gravitational field strength), this plot shows the shape of $\rho_{\text{fields}}$ becoming deformed in the strong gravity (low $M_{Pl}$) case. Remaining parameters were fixed at $f_a=10^{18}GeV$, $m_a=10^{-11}eV$, $q=1$, $q_m=4\pi$, $\alpha=1/4\pi$, $M_m=10^{22}GeV$.}
    \label{fig: gravvar}
\end{figure}

Finally, we can also isolate the purely gravitational effects and study the backreaction we expect from the highly coupled Einstein equations. In order to do this, we vary $M_{Pl}$ in Fig. \ref{fig: gravvar} to produce the effect of varying gravitational field strength without changing the intrinsic properties of either the monopole or the domain wall. In higher dimensional theories \cite{arkani-hamed_hierarchy_1998}, quantum gravity \cite{dvali_black_2010}, as well as modified gravity \cite{brans_machs_1961}, scale dependency or the running of $M_{Pl}$ can be found, which may justify our method of pinpointing gravitational behaviours thus. We observe that the $\rho_{\text{fields}}$ profile begins to deform at $M_{Pl}$ not much below the standard value of $\simeq10^{19}GeV$. This change in shape from the uniformly increasing, identical profiles seen at large $M_{Pl}$ (i.e. weak gravity) is evidence of non-trivial backreaction: as $M_{Pl}$ becomes comparable to the total mass, $C\simeq1$ everywhere no longer applies. Recasting the equation of motion (\ref{eq: Leom}) with this in mind, we find that the d'Alembertian is
\begin{equation}
\label{eq: gravboxa}
\begin{aligned}
    \Box a(r)&=g^{\mu\nu}\nabla_{\mu}\nabla_{\nu}a(r)\equiv\frac{1}{\sqrt{-g}}\frac{\partial}{\partial x^{\mu}}\left(\sqrt{-g}g^{\mu\nu}\left(\frac{\partial}{\partial x^{\nu}}a(r)\right)\right)\\
    &=C(r)a''(r)+C'(r)a'(r)+\left(\frac{2}{r}+\frac{N'(r)}{2N(r)}\right)C(r)a'(r),
\end{aligned}
\end{equation}
such that there is now a non-trivial $C'$ that can be interpreted heuristically as an effective ``gravitational friction" term, characterising an intricate dependence of $a$ upon the spacetime geometry and vice versa. This manifests as a change in $\rho_{\text{fields}}$ in Fig. \ref{fig: gravvar}, which in turn translates to the $a'$ modifications we previously found.

We conclude that given a model minimally coupling gravity to a Lagrangian describing an axion domain wall+monopole setup, it would be possible to recover a solitonic object obeying predetermined, physical boundary conditions for QCD parameters. Expanding the parameter space to include ALP models, we also found solutions in which horizons may form, for which our simple model employing matching conditions at the monopole core is a candidate regularised construction. From our construction, the monopole core boundary was taken to approximate a hypersurface across which our results would indicate there to be a redshift introduced by the timeshift corresponding to the value of $N(r_{\text{core}})=N_0 = \text{constant}$. Thus, a monopole bag may be a soliton as predicted by the flat space solution in \cite{kogan_axions_1993}, or appear to form a dyonic black hole-like object that may possess a non-singular centre. In either case, we discovered that the inclusion of gravitational considerations resulted in a deformation of the axion domain wall shape, leading to a compact object which could be equipped with an external axion profile, even if horizons develop.

We shall now make a brief statement about the stability of the solution, beginning with the central region. As demonstrated in the analysis of \cite{arreaga_stability_2000}, monopole stability depends on the values of charge, mass, and symmetry breaking scale. The relative magnitudes of these parameters give rise to a range of final states: a completely static monopole with fixed $r_{\text{core}}$, an `oscillating' solution with an exterior black hole, a bounce solution analogous to topological inflation as proposed by Barriola and Vilenkin \cite{barriola_gravitational_1989, vilenkin_topological_1994}, and no monotonically inflating or collapsing solutions for the motion of the monopole shell $r_{\text{core}}$. Reflecting on our findings, our extraneous mass-energy from the axion domain wall leads to an external approximate RN spacetime, which one may speculate to receive support from the shape of the axion potential and the effective negative cosmological constant in the monopole. Thus, if a regular centre could be found, given the model-dependence of solutions (the freedom to choose the symmetry breaking underlying the particular monopole, taking different couplings through choice of ALP model, etc.) whether the central region hidden behind horizons may be undergoing some exotic dynamics is another matter.

As for the linear and perturbative stability of the axion domain wall, several existing studies apply to our current scenario. As noted in §\ref{sec: mbag}, dissipative processes similar to those in flat space can be reasonably expected in the gravitating scenario, and can qualitatively motivate some expectation of longevity of the system though a rigorous stability proof remains to be performed in future. Certainly, the formation of horizons could alter the precise contents and nature of the emissions, but the (virtual) light particle pair creation described previously need not destroy or greatly compromise the monopole bag's stability given the existence of analogous bounds to the flat space case. In addition to decay channels available, in the most gravitationally relevant case, we observed partial regions of the axion profile existing outside of the horizons formed, thus we may reasonably wish to confirm the stability of this external wall. In \cite{rajaraman_solitons_1984}, linear perturbations were considered, and were not found to give rise to exponential growth in modes. Relatedly, \cite{blau_dynamics_1987, guven_covariant_1993, garriga_perturbations_1991} demonstrated that in domain walls and vacuum bubbles, while dynamics may prove non-trivial, perturbative stability is predicted. We acknowledge, however, that there remain certain possibilities not covered by present literature. For instance, a full perturbative analysis including gravitational and quantum effects is yet to be performed to our knowledge, in particular for theories with an axionic interaction term. Additionally, the consequences of dynamics such as rotational degrees of freedom and overall shape deformation are yet to be examined in the context of defect stability. These form future directions necessary for producing a complete dynamical model of our construction and similar objects. 


\section{Physical and Theoretical Implications}
\label{sec: physth}

The properties of non-standard black hole-like objects have long been a subject of interest, and there is much existing discussion relating to Hawking radiation in regularised spacetimes \cite{sueto_evaporation_2023, davies_nonsingular_2025}, the instability discovered at the Cauchy horizon \cite{poisson_internal_1990, ori_inner_1991} (an overview of related instabilities can be found in \cite{carballo-rubio_towards_2025}), and hairy black holes \cite{coleman_quantum_1992, bai_hairy_2021, visser_dirty_1992}. The system we consider appears to live at the intersection of these topics; as our simple model seems to point towards a class of black hole-type solutions, we will devote this section to discussing their projected evolution, then consider some of the late-time implications of such novel early universe objects in both the solitonic and black hole-like limit.

\subsection{Expected Evolution and Dynamics}
\label{sec: eer}

While our present analysis studies the possibility of a static final state, it may be interesting for us to speculate on the dynamics the gravitating monopole bag may undergo. As demonstrated in---for instance---\cite{widrow_dynamics_1989, gangui_topological_2001}, the removal of standard domain walls by dissipation occurs when the energy gradient or curvature from gravitational effects is sufficient to unwind the field $a(r)$. In this process, the radial configuration for the field oscillates and bounces until a flat, trivial profile is reached, either via complete dissipation or gravitational collapse resulting in a singular black hole. In the monopole bag system, the monopole electric charge is topologically protected by the nature of the monopole, and the axion vacuum is related to the remaining discrete symmetry. Despite distinct origins, the strong EM fields in the central region require that the two effects coincide, so that the topological structure at the centre is determined by the associated Witten effect.\footnote{Though higher order corrections to the coupling and linearity of the $aF\tilde{F}$ term may be expected, the discretised, topological nature of the two quantities protects them against such corrections and permits the close tuning necessary for this alignment in the core region.} As discussed, any tendency towards a variation in the value of $a(r\rightarrow0)$ away from $2\pi f_a$ incurs a great cost in electromagnetic energy density. This implies that in place of a trivialisation of the axion domain wall, one might expect the opposing EM and gravitational influences to balance, which was confirmed in §\ref{sec: sols}. However, some dissipation of the wall could still be anticipated even if a stable object can be sustained through the interactions supplied by the monopole bag setup. In addition to the $W^{\pm}$ pair creation already outlined, it seems that even if two horizons form as in Fig. \ref{fig: strong}, we can observe an object with an external axion domain wall which may emit some axions via higher order microscopic processes, though the dominant loss in such a case would be via Hawking radiation. Through Hawking radiation, one may expect the horizons to begin to coalesce, tending towards an extremal configuration; as the source of magnetic and electric charge both is the central monopole and the associated Witten effect, only mass will fall during radiation, which permits extremality to be attained. It can therefore be stipulated that our analysis of the gravitating monopole bag indicates that structures with similar features may have the potential to tend towards an extremal configuration should horizons form. We draw attention to a relevant and interesting description of magnetic/dyonic black hole evolution in \cite{bai_hairy_2021}. 

To fully assess expected evolution and for the sake of completeness we must address the instabilities accompanying RN and Kerr solutions possessing a Cauchy horizon. It was proposed by Penrose \cite{cecile_m_dewitt_battelle_1968} that the causal structure of spacetime within a Cauchy horizon renders the interior unphysical, in line with the strong cosmic censorship conjecture \cite{gnedin_destruction_1993}. It was argued that observers at the Cauchy horizon should encounter infinitely blueshifted radiation influx, which induces an exponentially growing mass function at the Cauchy horizon (a ``mass-inflation instability") and in turn results in a curvature singularity holding an unphysical spacetime within. As many regularised black hole spacetimes possess Cauchy horizons, the question of whether a similar instability could occur naturally arises. While there is not yet a strong consensus on the presence of a Cauchy instability in regularised objects with two horizons in literature\footnote{We direct the interested reader to \cite{poisson_internal_1990, bonanno_cauchy_2025, lee_topological_1987, maeda_novel_2005}.}, it is meaningful to note a particular proposal for the final state of objects with such an instability. Other than models that work to avoid the formation of a Cauchy horizon altogether, it has been proposed in \cite{filippo_fully_2024} that objects with extremal horizons constitute the most viable final stable states to black hole evolution by virtue of their independence of various classical and semi-classical instabilities, including mass inflation \cite{gajic_interior_2019}. Our construction thus far seems to agree that by standard evolution arguments and stability analysis both, should a stable regularised dyonic black hole be possible through a setup such as a monopole bag, an extremal black hole would seem to be the most promising candidate for a final state.  In our axion-monopole system, the central dyon charge is completely screened, which is justified by the equivalence between a dyon and an uncharged monopole under the corresponding axion vacuum values (as explained in §\ref{sec: eft}) originating from the periodicity of the axion field. We can therefore treat the actual charge of the monopole-bag system as localised on the domain wall, so that domain wall mass and the electrical charge of the monopole bag system have the same origin: the spatial derivative of the axion profile. This ``geometrical origin" of the spatial axion distribution is well-defined in our setup, and may correspond to regularisation in a thin-shell domain wall scenario. In this sense, if we imagine a domain wall-induced black hole dynamically transforming into a stable state due to the Cauchy instability, it could be natural to guess that the stable state is an extremal state. 


\subsection{Remnants and Late-time Consequences}

It is evident that our scenario shares many similarities with those predicting PBHs. In present literature, there are a significant number of models proposing that ALP parameters are able to facilitate domain wall formation during inflation due to allowing an $f_a$ above the usual range $10^8-10^{12}GeV$ allowed by observations in QCD contexts \cite{sato_unified_2018, gelmini_primordial_2023}. This causes them to be expanded outside of the horizon by inflation, leading to more massive walls that then re-enter after inflation. These closed walls have the potential to reach masses sufficient for catastrophic collapse, which is a cosmologically founded motivation for the study of ALP models in existing literature as well as the present study; there is much speculation on the parameter space thus formed PBHs may cover and the associated dark matter fraction. As has been shown extensively \cite{widrow_collapse_1989, silveira_dynamics_1988, vilenkin_cosmic_1985, huang_structure_1985}, the natural evolution of (axion) domain walls is an overall collapse due to intrinsic tension. The \textit{minimum} condition for PBH formation from this process is often stated as requiring the thickness $\delta$ of the closed wall to be smaller than the Schwarzschild radius $r_{\text{Sch}}$, since the wall becomes dynamical at $\delta$ in the sense that energy emission becomes significant \cite{ge_sublunar-mass_2020, vachaspati_lunar_2018, fort_global_1993, dunsky_primordial_2024}. Therefore, in requiring $\delta<r_{\text{Sch}}$, one demands that the total mass remains sufficiently constant so that the radius of the contracting wall can eventually shrink to within its Schwarzschild radius, thus concentrating enough mass-energy to initiate catastrophic collapse and form a black hole.

Recall that in our monopole bag model, we have not supposed the total mass to be constrained in any way. However, should we follow the formation pathway dictated by ALP domain walls synthesised during inflation, we can make a very na\"{i}ve estimate of the parameters usually required.\footnote{Notice that as Kogan pointed out \cite{kogan_axions_1993}, the monopole bag can also be formed through monopole-domain wall collisions, which could provide alternative initial properties for the monopole-axion domain wall system.} Suppose we assume that PBH formation demands $\delta<r_{\text{Sch}}$, i.e. $1/m_a<4\pi\sigma R^2/M_{Pl}^2$ where $\sigma$ is the wall's surface energy density and $R\gg \delta$ is its size at re-entry. Taking an arbitrary, conservative value of $R=10^{15}GeV^{-1}$ and using $\sigma\sim m_af_a^2$ \cite{vachaspati_lunar_2018, huang_structure_1985}, this condition translates to $m_af_a\gtrsim M_{Pl}/R\simeq10^{4}(GeV^2)$. Thus, one may expect $m_a=10^{-20}GeV$, $f_a=10^{18}GeV$ to not be able to reliably form PBHs. Certainly, many intricacies were not captured by such a rough order of magnitude estimate, however we simply wish to highlight that even without a realistic total mass conservation constraint, we were able to identify a scenario within our monopole bag framework that could potentially produce a black hole-like object by virtue of the axion-monopole interactions in the central region. At this selection of fairly moderate parameters that may otherwise not guarantee PBH formation in a pure domain wall model, the anchoring of the domain wall by the inner boundary conditions courtesy of the monopole bag construction may prevent complete dissipation. Further collapse could then perhaps occur, which one may conjecture to lead to the formation of a dyonic regular black hole; whether this is feasible for the monopole bag would demand a more careful analysis of the object's dynamics and a more pointed choice of a physical parameter space. Thus, specific quantitative estimates for the exact parameter spaces of interest and subsequent abundances are beyond the scope of this study, though our work aims to draw attention to the possibility of alternative structures expanding the range of PBH properties expected.

Furthermore, it is known that factors such as asphericity and angular momentum in domain walls may impact both the formation and the final properties of PBHs. The evolution of aspherical perturbations was first shown by Widrow \cite{widrow_collapse_1989}, who demonstrated that these perturbations are expected to be amplified in the later stages of a closed wall's collapse. This is significant for our scenario, in which a domain wall may be maintained throughout late-time evolution, making it likely that any asphericities can continue to impact the object's shape, interactions, and emissions beyond usual expectations. The latter is particularly interesting, as perfect spherical symmetry forbids the production of both gravitational and EM radiation. Novel constructions such as ours may represent exceptions where the expected signal can be estimated, contributing to (non-)detection bounds in future observations \cite{ito_gravitational_2024}.

Not only PBHs and non-singular black holes, but magnetic monopoles and their gravitating variants have also long been considered possible dark matter candidates \cite{burdyuzha_magnetic_2018, department_of_physics_and_astronomy_university_of_california_irvine_california_92697_usa_magnetic_2020, maldacena_comments_2021, diamond_constraints_2022, pazameta_general_2012, bai_primordial_2020}. There are many bounds on these candidates that deserve to be summarised. The most well-known constraint is the Parker bound \cite{turner_magnetic_1982} limiting the number density of magnetically charged species by their observable backreaction upon intergalactic magnetic fields. Another crucial consideration is that dark matter candidate objects cluster sufficiently around galaxies. For a magnetically charged species, it must not be accelerated to escape velocity by galactic magnetic fields in a timescale shorter than the time the galaxy has spent at its present background field strength \cite{perri_magnetic_2024}. Finally, it is necessary to consider the conditions during the object's formation. Specifically, as charged objects are accelerated by primordial fields, energy is lost from these primordial fields, therefore the abundance of charged species must not be such that these primordial fields cannot survive until present day. 

Detailed quantitative results can be found in \cite{turner_magnetic_1982, parker_magnetic_1987, parker_origin_1970, perri_magnetic_2024}, thus given the model-dependent nature of such analysis, it suffices at present to point out several assumptions in present literature that may require consolidation. Firstly, in determining these bounds, the loss of monopole's energy in interactions with the background medium is assumed negligible. For a compact, effective dyonic object, the scattering and momentum transfer estimates leading to the above conclusion should hold sufficiently to justify this assumption. However, any more spatially extensive object (such as those which are solitonic or black hole-like with an external axion profile) would warrant a reconsideration of the dissipative contributions. In particular, at early times, one may expect there to be a period during which our object is burdened with an axionic ``halo" whose frictional interaction with the surrounding (likely radiation dominated) medium must be analysed, demanding a more careful derivation of primordial field constraints. In our analysis, we have ignored all gauge field couplings with matter other than the axion. This allowed us to analyse the closed form within the three-field system of axion, EM, and gravity. On the other hand, since gravity universally couples with all matter and gauge fields, and the EM fields would couple with charged matter; when field strength exceeds the threshold energy, we may not be able to ignore the vacuum cloud or core formation of such fields. This may require extensions to our setup to capture these effects and describe reality. Such work would share the same essence of the core formation with a monopole in e.g. a flat GUT context \cite{preskill_magnetic_1984}, for which an analysis with gravity is left as a future interest. 

There is one further possibility that we find relevant to our scenario: the formation of an electroweak corona about magnetic black holes is a phenomenon predicted to have diverse effects on interactions with the intergalactic medium and Hawking radiation. When the magnetic field exceeds the electroweak scale, or in other words, when the event horizon lies within the scale $m_W^{-1}$ determined by $W$ mass, one may expect an external region of restored electroweak symmetry \cite{maldacena_comments_2021, bai_phenomenology_2020}. Directly quoting the result of \cite{bai_hairy_2021}, should our object of $q_m=4\pi/e, g=1$ have a mass below $9.3\times10^{35}GeV$, a so-called electroweak corona can be present. We previously observed that the inner horizon in e.g. Fig. \ref{fig: strong} has fallen within range of the monopole core size. This was our indication that for parameters considered so far one may indeed expect regions of restored electroweak symmetry as horizons sink within the monopole core. In such a case, Hawking radiation should be significantly accelerated, and interactions with ordinary matter would be modified. This would include the emission of additional, unique modes \cite{maldacena_comments_2021}, as well as the need for a careful reconsideration of clustering as a result of non-trivial plasma/interstellar medium interactions mediating phenomena including mergers and captures.


\subsection{No-Hair Conjecture}
\label{sec: hair}

While it is not the focus of our research, it would be prudent to succinctly address the no-hair conjecture and the position of our results within its framework. The no-hair conjecture originates from the fact that black holes should be uniquely specified by only their mass, charge, and angular momentum, while any additional information about the matter content of a black hole is not externally observable \cite{misner_gravitation_2017}. The conjecture has been scrutinised since the early days of its formulation, and many exceptions have since been raised (e.g. \cite{gao_black_2022}). In particular, a construction of our type appears to offer an interesting response to the Hui-Nicolis no-hair conjecture \cite{hui_no-hair_2013}.

We have obtained solutions possessing a spatially varying scalar field outside of the horizon, however the scalar field originates from and is anchored by the $aF\tilde{F}$ interaction term, and is not itself a conserved quantity in the theory, thereby making it non-essential and permitted. That is to say, should the coupling be switched off, the scalar field would vanish instantly, since the potential term alone would not give rise to the black hole-like solution we obtained; while indeed the potential alone would be sufficient to produce a domain wall, the monopole would not be subject to the Witten effect, and thus the system would be trivially decoupled. In fact, one may observe a manifestation of the principles underlying the no-hair conjecture in our setup as follows. Should one attempt to solve directly for $a'$, one could take the gravitating version of the equation of motion (\ref{eq: Leom}) (i.e. applying (\ref{eq: gravboxa})),
such that the analytical solution for $a$ may be found via the equation of motion
\begin{equation}
\label{eq: Ca'sol}
    \frac{1}{\sqrt{N}r^2}\frac{\partial}{\partial r}\left(\sqrt{N}r^2 Ca'\right)-\frac{\partial}{\partial a}V(a)+\frac{\alpha}{8\pi}\frac{1}{f_a}F_{\mu\nu}\tilde{F}^{\mu\nu}=0
\end{equation}
\begin{equation}
\label{eq: Ca'intsol}
    \implies a'=-\frac{1}{C}\frac{1}{4\pi}\frac{1}{\sqrt{N}r^2}\int^{\infty}_r\sqrt{N}r^2drd\Omega\left(\frac{\partial}{\partial a}V(a)-\frac{\alpha}{8\pi f_a}F_{\mu\nu}\tilde{F}^{\mu\nu}\right).
\end{equation}
At the horizon(s), $C=0$, which seems to lead to an unphysical divergent gradient. Certainly, if the model was such that $V(a)\geq0$ with no CS term, there would be no hope of recovering a finite term should a non-trivial $a$ profile occur at the horizon. This would concur with the no-hair arguments made in, for instance, \cite{bekenstein_novel_1995}. However, the presence of an axion potential $V(a)$ as well as a CS term permits the cancellation necessary to circumvent the no-hair conjecture. At the horizon, though $Ca'=0$, $C'$ is finite, and the system described by the equation of motion in (\ref{eq: Ca'sol}) is physical given the integral on the right-hand side of (\ref{eq: Ca'intsol}) must sum to zero. The former statement implies a level of $a'$ tuning at the horizon is required in order to permit the survival of the scalar field. This cannot be achieved without the cancellation of summed EM and potential terms, which is in turn made possible by the formulation of an axion domain wall.\footnote{It is also by this logic that we were able to numerically deal with cases where horizons were formed. While strictly, regions in which $C<0$ are physically ambiguous as the application of the usual differential equations may not be valid, given our focus on the behaviour of the axion field, we follow a method common in literature (e.g. \cite{lee_charged_1991}) where the extension to $C<0$ regions in numerical solutions is done simply by applying static local boundary conditions.}


\section{Summary and Outlook}
\label{sec: sum}

We begin by introducing the monopole bag, an object constructed by placing a monopole in a closed axion domain wall. In flat space, it has an energetically preferred solitonic stable state---the ``monopole bag". This formulation, however, does not consider the gravitational properties of the system, which one may expect to become significant at large values of the axion decay constant allowed by ALP models. We thus proceeded to construct a gravitating monopole bag, finding that a stable, dyonic, black hole-like state can also result for input parameters in the range of ALP models. By virtue of the matching conditions imposed and the Witten effect, we hypothesised that the core of our state might be regularisable, which would permit it to evade the Bronnikov condition \cite{bronnikov_regular_2001}, a no-go theorem typically excluding the existence of regular black holes with electric charge. This is a property distinguishing our solution from standard magnetised regular black hole results. We also discovered that the shape of the axion domain wall can be significantly distorted by gravity; even with the formation of horizons, this profile may extend outside the black hole in our simple model. We emphasise at this stage, however, that a rigorous black hole construction of this type---and thus a strong conclusion as to the nature of the final state---would necessitate dynamical evolution, a detailed description of the core region, as well as ensuing discussions of perturbative stability. This is beyond the scope of our present analysis.

The object we have proposed is a prototype for a future novel construction of a dyonic black hole with a regularised centre. By investigating the structure of the static case, our results suggest a possible track to achieve this highly sought-after state without the need for exotic gravitational (such as particular modified gravity models or non-standard fluid properties \cite{ayon-beato_bardeen_2000, junior_dyonic_2025}) or electromagnetic behaviours (such as complex dark sector phenomenology and higher order terms \cite{felice_exotic_2025, bronnikov_regular_2001}). Regularity of the centre could be justified by the physically well-founded properties of magnetic monopoles, and the final remnant presents an interest both as a theoretical object shedding light on the no-hair conjecture and at late-times as a dark matter component. Particularly for the latter, Planck-scale objects have been highlighted as promising for bettering theoretical and phenomenological understanding \cite{profumo_probing_2026}; we expect that a rigorous construction with such motivations and drawing inspiration from EM effects studied in this work may prove a fruitful future direction. Furthermore, we have illustrated that compact objects suffer non-trivial corrections to their shape and thus energy distribution from gravitational effects, which is a sign that careful analysis of their dynamics and interactions with the galactic medium may prove essential to understanding their properties and stability. The results may also prove applicable to analogous scenarios such as other shell-like structures (e.g. bubbles, overdensity regions, and gravastars). Our construction is also motivated by continued efforts to find a pleasing solution to the information paradox; the possibility of black hole evaporation producing a non-singular final state via reaching extremality \cite{sueto_evaporation_2023} is a decidedly simple and thus attractive option. Additionally, axion-monopole interactions such as those discussed in \cite{co_dark_2025} also present interesting possibilities, for the likes of which a particular construction such as ours may provide a novel perspective.

There are several future directions that we would like to highlight. Firstly, the non-trivial axion profile introduced could offer a new angle in considering the dynamical properties of a system beyond our current static ansatz. Since the profile possesses a CP-violating background, dynamical aspects of the axion domain wall may simultaneously realise CP violation and a non-equilibrium environment. This has the potential to fit the Sakharov conditions for baryon asymmetry generation, which encourages an investigation of the dynamical aspects of the monopole-axion domain wall system in light of this long-standing mystery. Notably, gravitation might not respect global symmetry preservation, which might relate to the quality problem. It is known that the quality of PQ symmetry and CP violation has an impact on the axion vacuum value in the original axion potential, which in turn relates to the existence of a bias potential in the domain wall formalism \cite{murai_domain_2024}. While we have assumed a high quality axion, it should be noted that the axion's quality may also impact the uniformity of its field value through the mismatch between the complete screening of the central monopole charge and the asymptotic axion vacuum value determined by the original potential.\footnote{Notice that since the periodicity and equivalence of the axion vacuum and potential originate from the fact that the axionic degree of freedom is an angular variable or compact field space, this basic property may be preserved for any quality. Such topological properties may be stable against continuous deformation like quantum corrections or local small perturbations.}

In \cite{loladze_monopole-fermion_2024}, it was pointed out that a light charged particle interacting with a monopole is regarded as a dressed charged particle carrying a phase (``rotor") degree of freedom. It may originate from the Higgs degree of freedom giving it mass.\footnote{Note that the Higgs sector associated with the phase does not always coincide with the monopole Higgs sector. Additionally, charged fermions with no such Higgs coupling (``massless charged fermions") cannot be so dressed. This would lead to the non-trivial scattering behaviours tied to the well-known Callan-Rubakov effect \cite{callan_dyon-fermion_1982, rubakov_adler-bell-jackiw_1982}.} The adiabatic (continuous) dressed phase shift (or the corresponding ``physical" gauge transformation phase shift) in the dressed charged particle may resolve the long-standing missing final state (unitarity) puzzle. This may yield free propagation of a charged particle (such as an electron) on the central monopole by cancelling the effect of the spin dependent monopole magnetic force with a geometrical force and also by enhancing the helicity rotation in the mass term, analogous to a Berry phase gauge boson (or Higgs kinetic pressure) which prevents anomalous absorption at the central monopole core. The shift of the phase position is a zero mode moduli, on the other hand, the electron may feel the effects as a backreacted Higgs moduli dynamics along with electron trajectory, which could backreact to its trajectory in turn. It does not necessarily excite a quantised constant rotational velocity for the phase degree of freedom sometimes called a rotor, which would correspond to a gapped excitation of a dyon.\footnote{The corresponding rotor to excite a charge for the central monopole would have to be one associated with a rotor in the monopole Higgs sector, not those of other Higgs sectors. The high cost rotor can realise highly localised charge within the monopole core.} As already indicated, such a gapped electric charge transition is what permits our monopole bag system to be stable. This new viewpoint might also encourage a fresh analysis of bound state or core formation with a central monopole, as well as the efficiency of UV processes (e.g. GUT processes) in the core region. \footnote{We clarify at this stage the distinction between dyonic states arising from internal space dynamics described by the local field theory, and the static dyonic charge that can be induced via the Witten effect in an axionic model. Thus, the application of rotors demands some caution and future scrutiny.} Furthermore, charged and magnetised objects may leave detectable traces or even catalyse exotic processes in the cosmic medium at various stages of cosmic evolution, for example, following a similar line of inquiry to studies of heavy charged particles \cite{dubrovich_primordial_2004, kohri_big_2007}. An estimate of the extent and specific nature of these consequences will be left for future studies.

A slightly different direction for further investigation would be to consider a more general class of regular magnetic black holes while taking advantage of the axionic interactions, such as in the context of non-linear electrodynamics (NLED). In \cite{felice_exotic_2025}, a compact object was constructed by permitting higher order EM interaction terms in the theory. Similarly, we expect that an extension with an additional $f(a)F\tilde{F}$ term (where $f$ is some function of the axion field $a$ allowing shift symmetry and potential periodicity) may offer new insight on instability discussions intrinsic to such constructions, and thus perhaps contribute another possibility for an exotic compact object. 

Finally, a more rigorous formulation of the monopole's core structure would be necessary for determining not only the stability of the object up to quantum and perturbative effects, but also the decay channels open to the construction. One way of understanding the behaviour of the core in the presence of an axion profile would be to search for a stable configuration within the full theory including the central non-Abelian region, which might be illuminated by the well-known analyses of the moduli space of monopoles \cite{donaldson_nahms_1984, harvey_magnetic_1996}. While this particular example applies only to BPS states, we suggest that this leads naturally to our interest in understanding the UV foundations of our setup and beginning to formulate a more complete model. Within the core region, approximating BPS, identifying deviations from this limit, and considering axion dynamics could be a natural framework with which to study the highly compact limit introduced in Fig. \ref{fig: smallMm}. The relaxation of the axion-monopole system might be investigated thus; this would aid in clarifying the possibility of maintaining regularity despite collapse, and the decays that may compromise it should stability be attainable. One may envisage this to also be tied to our interests in NLED, which might be expected to come into play at small radii within our current construction.


\newpage

\printbibliography

@article{ito_gravitational_2024,
	title = {Gravitational {Wave} {Search} through {Electromagnetic} {Telescopes}},
	volume = {2024},
	copyright = {https://creativecommons.org/licenses/by/4.0/},
	issn = {2050-3911},
	url = {https://academic.oup.com/ptep/article/doi/10.1093/ptep/ptae004/7515273},
	doi = {10.1093/ptep/ptae004},
	language = {en},
	number = {2},
	urldate = {2026-04-27},
	journal = {Progress of Theoretical and Experimental Physics},
	author = {Ito, Asuka and Kohri, Kazunori and Nakayama, Kazunori},
	month = feb,
	year = {2024},
	pages = {023E03},
}

@article{kohri_big_2007,
	title = {Big bang nucleosynthesis with long-lived charged massive particles},
	volume = {76},
	copyright = {http://link.aps.org/licenses/aps-default-license},
	issn = {1550-7998, 1550-2368},
	url = {https://link.aps.org/doi/10.1103/PhysRevD.76.063507},
	doi = {10.1103/PhysRevD.76.063507},
	language = {en},
	number = {6},
	urldate = {2026-04-27},
	journal = {Physical Review D},
	author = {Kohri, Kazunori and Takayama, Fumihiro},
	month = sep,
	year = {2007},
	pages = {063507},
}

@article{agrawal_monodromic_2024,
	title = {The monodromic axion-photon coupling},
	volume = {2024},
	issn = {1029-8479},
	url = {https://link.springer.com/10.1007/JHEP01(2024)169},
	doi = {10.1007/JHEP01(2024)169},
	language = {en},
	number = {1},
	urldate = {2026-04-27},
	journal = {Journal of High Energy Physics},
	author = {Agrawal, Prateek and Platschorre, Arthur},
	month = jan,
	year = {2024},
	pages = {169},
}

@misc{cirelli_dark_2025,
	title = {Dark {Matter}},
	url = {http://arxiv.org/abs/2406.01705},
	doi = {10.48550/arXiv.2406.01705},
	urldate = {2026-04-27},
	publisher = {arXiv},
	author = {Cirelli, Marco and Strumia, Alessandro and Zupan, Jure},
	month = dec,
	year = {2025},
	note = {arXiv:2406.01705 [hep-ph]},
}

@misc{murai_domain_2024,
	title = {Domain walls in {Nelson}-{Barr} axion model},
	url = {http://arxiv.org/abs/2412.19456},
	doi = {10.48550/arXiv.2412.19456},
	urldate = {2026-04-27},
	publisher = {arXiv},
	author = {Murai, Kai and Nakayama, Kazunori},
	month = dec,
	year = {2024},
	note = {arXiv:2412.19456 [hep-ph]},
}

@article{dubrovich_primordial_2004,
	title = {Primordial bound systems of superheavy particles as the source of ultra-high energy cosmic rays},
	volume = {22},
	copyright = {https://www.elsevier.com/tdm/userlicense/1.0/},
	issn = {09276505},
	url = {https://linkinghub.elsevier.com/retrieve/pii/S0927650504001252},
	doi = {10.1016/j.astropartphys.2004.07.002},
	language = {en},
	number = {2},
	urldate = {2026-04-27},
	journal = {Astroparticle Physics},
	author = {Dubrovich, V.K. and Fargion, D. and Khlopov, M.Yu.},
	month = nov,
	year = {2004},
	pages = {183--197},
}

@article{gajic_interior_2019,
	title = {The interior of dynamical extremal black holes in spherical symmetry},
	volume = {1},
	issn = {2578-5885, 2578-5893},
	url = {http://arxiv.org/abs/1709.09137},
	doi = {10.2140/paa.2019.1.263},
	number = {2},
	urldate = {2026-04-25},
	journal = {Pure and Applied Analysis},
	author = {Gajic, Dejan and Luk, Jonathan},
	month = apr,
	year = {2019},
	note = {arXiv:1709.09137 [gr-qc]},
	pages = {263--326},
}

@misc{ansoldi_spherical_2008,
	title = {Spherical black holes with regular center: a review of existing models including a recent realization with {Gaussian} sources},
	shorttitle = {Spherical black holes with regular center},
	url = {http://arxiv.org/abs/0802.0330},
	doi = {10.48550/arXiv.0802.0330},
	urldate = {2026-04-22},
	publisher = {arXiv},
	author = {Ansoldi, Stefano},
	month = feb,
	year = {2008},
	note = {arXiv:0802.0330 [gr-qc]},
}

@article{chen_black_2015,
	title = {Black {Hole} {Remnants} and the {Information} {Loss} {Paradox}},
	volume = {603},
	issn = {03701573},
	url = {http://arxiv.org/abs/1412.8366},
	doi = {10.1016/j.physrep.2015.10.007},
	urldate = {2026-04-22},
	journal = {Physics Reports},
	author = {Chen, Pisin and Ong, Yen Chin and Yeom, Dong-han},
	month = nov,
	year = {2015},
	note = {arXiv:1412.8366 [gr-qc]},
	pages = {1--45},
}

@article{cardoso_testing_2019,
	title = {Testing the nature of dark compact objects: a status report},
	volume = {22},
	issn = {2367-3613, 1433-8351},
	shorttitle = {Testing the nature of dark compact objects},
	url = {http://arxiv.org/abs/1904.05363},
	doi = {10.1007/s41114-019-0020-4},
	number = {1},
	urldate = {2026-04-22},
	journal = {Living Reviews in Relativity},
	author = {Cardoso, Vitor and Pani, Paolo},
	month = dec,
	year = {2019},
	note = {arXiv:1904.05363 [gr-qc]},
	pages = {4},
}

@article{kibble_topology_1976,
	title = {Topology of cosmic domains and strings},
	volume = {9},
	issn = {0305-4470, 1361-6447},
	url = {https://iopscience.iop.org/article/10.1088/0305-4470/9/8/029},
	doi = {10.1088/0305-4470/9/8/029},
	number = {8},
	urldate = {2026-04-22},
	journal = {Journal of Physics A: Mathematical and General},
	author = {Kibble, T W B},
	month = aug,
	year = {1976},
	pages = {1387--1398},
}

@article{green_primordial_2021,
	title = {Primordial {Black} {Holes} as a dark matter candidate},
	volume = {48},
	issn = {0954-3899, 1361-6471},
	url = {http://arxiv.org/abs/2007.10722},
	doi = {10.1088/1361-6471/abc534},
	number = {4},
	urldate = {2026-04-22},
	journal = {Journal of Physics G: Nuclear and Particle Physics},
	author = {Green, Anne M. and Kavanagh, Bradley J.},
	month = apr,
	year = {2021},
	note = {arXiv:2007.10722 [astro-ph]},
	pages = {043001},
}

@article{carr_black_1974,
	title = {Black {Holes} in the {Early} {Universe}},
	volume = {168},
	issn = {0035-8711, 1365-2966},
	url = {https://academic.oup.com/mnras/article-lookup/doi/10.1093/mnras/168.2.399},
	doi = {10.1093/mnras/168.2.399},
	language = {en},
	number = {2},
	urldate = {2026-04-22},
	journal = {Monthly Notices of the Royal Astronomical Society},
	author = {Carr, B. J. and Hawking, S. W.},
	month = aug,
	year = {1974},
	pages = {399--415},
}

@article{preskill_cosmology_1983,
	title = {Cosmology of the invisible axion},
	volume = {120},
	copyright = {https://www.elsevier.com/tdm/userlicense/1.0/},
	issn = {03702693},
	url = {https://linkinghub.elsevier.com/retrieve/pii/0370269383906378},
	doi = {10.1016/0370-2693(83)90637-8},
	language = {en},
	number = {1-3},
	urldate = {2026-04-22},
	journal = {Physics Letters B},
	author = {Preskill, John and Wise, Mark B. and Wilczek, Frank},
	month = jan,
	year = {1983},
	pages = {127--132},
}

@article{marsh_axion_2016,
	title = {Axion {Cosmology}},
	volume = {643},
	issn = {03701573},
	url = {http://arxiv.org/abs/1510.07633},
	doi = {10.1016/j.physrep.2016.06.005},
	urldate = {2026-04-22},
	journal = {Physics Reports},
	author = {Marsh, David J. E.},
	month = jul,
	year = {2016},
	note = {arXiv:1510.07633 [astro-ph]},
	pages = {1--79},
}

@article{brans_machs_1961,
	title = {Mach's {Principle} and a {Relativistic} {Theory} of {Gravitation}},
	volume = {124},
	copyright = {http://link.aps.org/licenses/aps-default-license},
	issn = {0031-899X},
	url = {https://link.aps.org/doi/10.1103/PhysRev.124.925},
	doi = {10.1103/PhysRev.124.925},
	language = {en},
	number = {3},
	urldate = {2026-04-22},
	journal = {Physical Review},
	author = {Brans, C. and Dicke, R. H.},
	month = nov,
	year = {1961},
	pages = {925--935},
}

@article{arkani-hamed_hierarchy_1998,
	title = {The {Hierarchy} {Problem} and {New} {Dimensions} at a {Millimeter}},
	volume = {429},
	issn = {03702693},
	url = {http://arxiv.org/abs/hep-ph/9803315},
	doi = {10.1016/S0370-2693(98)00466-3},
	number = {3-4},
	urldate = {2026-04-22},
	journal = {Physics Letters B},
	author = {Arkani-Hamed, Nima and Dimopoulos, Savas and Dvali, Gia},
	month = jun,
	year = {1998},
	note = {arXiv:hep-ph/9803315},
	pages = {263--272},
}

@article{dvali_black_2010,
	title = {Black {Holes} and {Large} {N} {Species} {Solution} to the {Hierarchy} {Problem}},
	volume = {58},
	issn = {0015-8208, 1521-3978},
	url = {http://arxiv.org/abs/0706.2050},
	doi = {10.1002/prop.201000009},
	number = {6},
	urldate = {2026-04-22},
	journal = {Fortschritte der Physik},
	author = {Dvali, Gia},
	month = jun,
	year = {2010},
	note = {arXiv:0706.2050 [hep-th]},
	pages = {528--536},
}

@article{silverstein_monodromy_2008,
	title = {Monodromy in the {CMB}: {Gravity} waves and string inflation},
	volume = {78},
	copyright = {http://link.aps.org/licenses/aps-default-license},
	issn = {1550-7998, 1550-2368},
	shorttitle = {Monodromy in the {CMB}},
	url = {https://link.aps.org/doi/10.1103/PhysRevD.78.106003},
	doi = {10.1103/PhysRevD.78.106003},
	language = {en},
	number = {10},
	urldate = {2026-04-21},
	journal = {Physical Review D},
	author = {Silverstein, Eva and Westphal, Alexander},
	month = nov,
	year = {2008},
	pages = {106003},
}

@article{bernal_alp_2021,
	title = {{ALP} {Dark} {Matter} in a {Primordial} {Black} {Hole} {Dominated} {Universe}},
	volume = {104},
	issn = {2470-0010, 2470-0029},
	url = {http://arxiv.org/abs/2110.04312},
	doi = {10.1103/PhysRevD.104.123536},
	number = {12},
	urldate = {2026-04-18},
	journal = {Physical Review D},
	author = {Bernal, Nicol\'as and Perez-Gonzalez, Yuber F. and Xu, Yong and Zapata, \'Oscar},
	month = dec,
	year = {2021},
	note = {arXiv:2110.04312 [hep-ph]},
	pages = {123536},
}

@article{navas_review_2024,
	title = {Review of {Particle} {Physics}},
	volume = {110},
	issn = {2470-0010, 2470-0029},
	url = {https://link.aps.org/doi/10.1103/PhysRevD.110.030001},
	doi = {10.1103/PhysRevD.110.030001},
	language = {en},
	number = {3},
	urldate = {2026-04-08},
	journal = {Physical Review D},
	author = {Navas, S. and Amsler, C. and Gutsche, T. and Hanhart, C. and Hernández-Rey, J.J. and Lourenço, C. and Masoni, A. and Mikhasenko, M. and Mitchell, R.E. and Patrignani, C. and Schwanda, C. and Spanier, S. and Venanzoni, G. and Yuan, C.Z. and Agashe, K. and Aielli, G. and Allanach, B.C. and Alvarez-Muñiz, J. and Antonelli, M. and Aschenauer, E.C. and Asner, D.M. and Assamagan, K. and Baer, H. and Banerjee, Sw. and Barnett, R.M. and Baudis, L. and Bauer, C.W. and Beatty, J.J. and Beringer, J. and Bettini, A. and Biebel, O. and Black, K.M. and Blucher, E. and Bonventre, R. and Briere, R.A. and Buckley, A. and Burkert, V.D. and Bychkov, M.A. and Cahn, R.N. and Cao, Z. and Carena, M. and Casarosa, G. and Ceccucci, A. and Cerri, A. and Chivukula, R.S. and Cowan, G. and Cranmer, K. and Crede, V. and Cremonesi, O. and D’Ambrosio, G. and Damour, T. and De Florian, D. and De Gouvêa, A. and DeGrand, T. and Demers, S. and Demiragli, Z. and Dobrescu, B.A. and D’Onofrio, M. and Doser, M. and Dreiner, H.K. and Eerola, P. and Egede, U. and Eidelman, S. and El-Khadra, A.X. and Ellis, J. and Eno, S.C. and Erler, J. and Ezhela, V.V. and Fava, A. and Fetscher, W. and Fields, B.D. and Freitas, A. and Gallagher, H. and Gershon, T. and Gershtein, Y. and Gherghetta, T. and Gonzalez-Garcia, M.C. and Goodman, M. and Grab, C. and Gritsan, A.V. and Grojean, C. and Groom, D.E. and Grünewald, M. and Gurtu, A. and Haber, H.E. and Hamel, M. and Hashimoto, S. and Hayato, Y. and Hebecker, A. and Heinemeyer, S. and Hikasa, K. and Hisano, J. and Höcker, A. and Holder, J. and Hsu, L. and Huston, J. and Hyodo, T. and Ianni, Al. and Kado, M. and Karliner, M. and Katz, U.F. and Kenzie, M. and Khoze, V.A. and Klein, S.R. and Krauss, F. and Kreps, M. and Križan, P. and Krusche, B. and Kwon, Y. and Lahav, O. and Lellouch, L.P. and Lesgourgues, J. and Liddle, A.R. and Ligeti, Z. and Lin, C.-J. and Lippmann, C. and Liss, T.M. and Lister, A. and Littenberg, L. and Lugovsky, K.S. and Lugovsky, S.B. and Lusiani, A. and Makida, Y. and Maltoni, F. and Manohar, A.V. and Marciano, W.J. and Matthews, J. and Meißner, U.-G. and Melzer-Pellmann, I.-A. and Mertsch, P. and Miller, D.J. and Milstead, D. and Mönig, K. and Molaro, P. and Moortgat, F. and Moskovic, M. and Nagata, N. and Nakamura, K. and Narain, M. and Nason, P. and Nelles, A. and Neubert, M. and Nir, Y. and O’Connell, H.B. and O’Hare, C.A.J. and Olive, K.A. and Peacock, J.A. and Pianori, E. and Pich, A. and Piepke, A. and Pietropaolo, F. and Pomarol, A. and Pordes, S. and Profumo, S. and Quadt, A. and Rabbertz, K. and Rademacker, J. and Raffelt, G. and Ramsey-Musolf, M. and Richardson, P. and Ringwald, A. and Robinson, D.J. and Roesler, S. and Rolli, S. and Romaniouk, A. and Rosenberg, L.J and Rosner, J.L. and Rybka, G. and Ryskin, M.G. and Ryutin, R.A. and Safdi, B. and Sakai, Y. and Sarkar, S. and Sauli, F. and Schneider, O. and Schönert, S. and Scholberg, K. and Schwartz, A.J. and Schwiening, J. and Scott, D. and Sefkow, F. and Seljak, U. and Sharma, V. and Sharpe, S.R. and Shiltsev, V. and Signorelli, G. and Silari, M. and Simon, F. and Sjöstrand, T. and Skands, P. and Skwarnicki, T. and Smoot, G.F. and Soffer, A. and Sozzi, M.S. and Spiering, C. and Stahl, A. and Sumino, Y. and Takahashi, F. and Tanabashi, M. and Tanaka, J. and Taševský, M. and Terao, K. and Terashi, K. and Terning, J. and Thoma, U. and Thorne, R.S. and Tiator, L. and Titov, M. and Tovey, D.R. and Trabelsi, K. and Urquijo, P. and Valencia, G. and Van De Water, R. and Varelas, N. and Verde, L. and Vivarelli, I. and Vogel, P. and Vogelsang, W. and Vorobyev, V. and Wakely, S.P. and Walkowiak, W. and Walter, C.W. and Wands, D. and Weinberg, D.H. and Weinberg, E.J. and Wermes, N. and White, M. and Wiencke, L.R. and Willocq, S. and Woody, C.L. and Workman, R.L. and Yao, W.-M. and Yokoyama, M. and Yoshida, R. and Zanderighi, G. and Zeller, G.P. and Zhu, R.-Y. and Zhu, S.-L. and Zimmermann, F. and Zyla, P.A. and Anderson, J. and Kramer, M. and Schaffner, P. and Zheng, W. and {Particle Data Group Collaboration}},
	month = aug,
	year = {2024},
	pages = {030001},
}

@article{bekenstein_novel_1995,
	title = {Novel ``no-scalar-hair" theorem for black holes},
	volume = {51},
	copyright = {http://link.aps.org/licenses/aps-default-license},
	issn = {0556-2821},
	url = {https://link.aps.org/doi/10.1103/PhysRevD.51.R6608},
	doi = {10.1103/PhysRevD.51.R6608},
	language = {en},
	number = {12},
	urldate = {2026-04-02},
	journal = {Physical Review D},
	author = {Bekenstein, Jacob D.},
	month = jun,
	year = {1995},
	pages = {R6608--R6611},
}

@article{sonoda_decay_1984,
	title = {Decay of a dyon},
	volume = {238},
	copyright = {https://www.elsevier.com/tdm/userlicense/1.0/},
	issn = {05503213},
	url = {https://linkinghub.elsevier.com/retrieve/pii/0550321384904498},
	doi = {10.1016/0550-3213(84)90449-8},
	language = {en},
	number = {2},
	urldate = {2026-03-29},
	journal = {Nuclear Physics B},
	author = {Sonoda, Hidenori},
	month = jun,
	year = {1984},
	pages = {259--276},
}

@article{rubakov_adler-bell-jackiw_1982,
	title = {Adler-{Bell}-{Jackiw} anomaly and fermion-number breaking in the presence of a magnetic monopole},
	volume = {203},
	copyright = {https://www.elsevier.com/tdm/userlicense/1.0/},
	issn = {05503213},
	url = {https://linkinghub.elsevier.com/retrieve/pii/0550321382900347},
	doi = {10.1016/0550-3213(82)90034-7},
	language = {en},
	number = {2},
	urldate = {2026-03-29},
	journal = {Nuclear Physics B},
	author = {Rubakov, V.A.},
	month = aug,
	year = {1982},
	pages = {311--348},
}

@article{callan_dyon-fermion_1982,
	title = {Dyon-fermion dynamics},
	volume = {26},
	copyright = {http://link.aps.org/licenses/aps-default-license},
	issn = {0556-2821},
	url = {https://link.aps.org/doi/10.1103/PhysRevD.26.2058},
	doi = {10.1103/PhysRevD.26.2058},
	language = {en},
	number = {8},
	urldate = {2026-03-29},
	journal = {Physical Review D},
	author = {Callan, Curtis G.},
	month = oct,
	year = {1982},
	pages = {2058--2068},
}

@article{gelmini_cosmology_1989,
	title = {Cosmology of biased discrete symmetry breaking},
	volume = {39},
	copyright = {http://link.aps.org/licenses/aps-default-license},
	issn = {0556-2821},
	url = {https://link.aps.org/doi/10.1103/PhysRevD.39.1558},
	doi = {10.1103/PhysRevD.39.1558},
	language = {en},
	number = {6},
	urldate = {2026-03-28},
	journal = {Physical Review D},
	author = {Gelmini, Graciela B. and Gleiser, Marcelo and Kolb, Edward W.},
	month = mar,
	year = {1989},
	pages = {1558--1566},
}

@article{filippo_fully_2024,
	title = {Fully extremal black holes: a black hole graveyard?},
	volume = {33},
	issn = {0218-2718, 1793-6594},
	shorttitle = {Fully extremal black holes},
	url = {http://arxiv.org/abs/2405.08069},
	doi = {10.1142/S0218271824400054},
	number = {15},
	urldate = {2026-03-26},
	journal = {International Journal of Modern Physics D},
	author = {Filippo, Francesco Di and Liberati, Stefano and Visser, Matt},
	month = nov,
	year = {2024},
	note = {arXiv:2405.08069 [gr-qc]},
	pages = {2440005},
}

@article{carballo-rubio_towards_2025,
	title = {Towards a {Non}-singular {Paradigm} of {Black} {Hole} {Physics}},
	volume = {2025},
	issn = {1475-7516},
	url = {http://arxiv.org/abs/2501.05505},
	doi = {10.1088/1475-7516/2025/05/003},
	number = {05},
	urldate = {2026-03-26},
	journal = {Journal of Cosmology and Astroparticle Physics},
	author = {Carballo-Rubio, Raúl and Filippo, Francesco Di and Liberati, Stefano and Visser, Matt and Arrechea, Julio and Barceló, Carlos and Bonanno, Alfio and Borissova, Johanna and Boyanov, Valentin and Cardoso, Vitor and Porro, Francesco Del and Eichhorn, Astrid and Jampolski, Daniel and Martín-Moruno, Prado and Mazza, Jacopo and McMaken, Tyler and Panassiti, Antonio and Pani, Paolo and Platania, Alessia and Rezzolla, Luciano and Vellucci, Vania},
	month = may,
	year = {2025},
	note = {arXiv:2501.05505 [gr-qc]},
	pages = {003},
}

@misc{co_dark_2025,
	title = {Dark {Matter} and {Baryon} {Asymmetry} from {Monopole}-{Axion} {Interactions}},
	url = {http://arxiv.org/abs/2511.10603},
	doi = {10.48550/arXiv.2511.10603},
	urldate = {2026-03-23},
	publisher = {arXiv},
	author = {Co, Raymond T. and Harigaya, Keisuke and Wang, Isaac R. and Xiao, Huangyu},
	month = nov,
	year = {2025},
	note = {arXiv:2511.10603 [hep-ph]},
}

@misc{gangui_topological_2001,
	title = {Topological {Defects} in {Cosmology}},
	url = {http://arxiv.org/abs/astro-ph/0110285},
	doi = {10.48550/arXiv.astro-ph/0110285},
	urldate = {2026-03-16},
	publisher = {arXiv},
	author = {Gangui, Alejandro},
	month = oct,
	year = {2001},
	note = {arXiv:astro-ph/0110285},
}

@article{gould_magnetic_2017,
	title = {Magnetic {Monopole} {Mass} {Bounds} from {Heavy}-{Ion} {Collisions} and {Neutron} {Stars}},
	volume = {119},
	copyright = {https://link.aps.org/licenses/aps-default-license},
	issn = {0031-9007, 1079-7114},
	url = {https://link.aps.org/doi/10.1103/PhysRevLett.119.241601},
	doi = {10.1103/PhysRevLett.119.241601},
	language = {en},
	number = {24},
	urldate = {2026-03-16},
	journal = {Physical Review Letters},
	author = {Gould, Oliver and Rajantie, Arttu},
	month = dec,
	year = {2017},
	pages = {241601},
}

@misc{profumo_probing_2026,
	title = {Probing {Planck}-{Scale} {Physics} with {High}-{Frequency} {Gravitational} {Waves}},
	url = {http://arxiv.org/abs/2603.02493},
	doi = {10.48550/arXiv.2603.02493},
	urldate = {2026-03-07},
	publisher = {arXiv},
	author = {Profumo, Stefano},
	month = mar,
	year = {2026},
	note = {arXiv:2603.02493 [hep-ph]},
}

@article{collaboration_planck_2020,
	title = {Planck 2018 results. {VI}. {Cosmological} parameters},
	volume = {641},
	issn = {0004-6361, 1432-0746},
	url = {http://arxiv.org/abs/1807.06209},
	doi = {10.1051/0004-6361/201833910},
	urldate = {2026-03-07},
	journal = {Astronomy \& Astrophysics},
	author = {Collaboration, Planck},
	month = sep,
	year = {2020},
	note = {arXiv:1807.06209 [astro-ph]},
	pages = {A6},
}

@article{collaboration_desi_2025,
	title = {{DESI} {DR2} {Results} {II}: {Measurements} of {Baryon} {Acoustic} {Oscillations} and {Cosmological} {Constraints}},
	volume = {112},
	issn = {2470-0010, 2470-0029},
	shorttitle = {{DESI} {DR2} {Results} {II}},
	url = {http://arxiv.org/abs/2503.14738},
	doi = {10.1103/tr6y-kpc6},
	number = {8},
	urldate = {2026-03-07},
	journal = {Physical Review D},
	author = {Collaboration, DESI},
	month = oct,
	year = {2025},
	note = {arXiv:2503.14738 [astro-ph]},
	pages = {083515},
}

@article{ferreira_bao-cmb_2026,
	title = {{BAO}-{CMB} tension and implications for inflation},
	volume = {113},
	issn = {2470-0010, 2470-0029},
	url = {https://link.aps.org/doi/10.1103/lq71-b84v},
	doi = {10.1103/lq71-b84v},
	language = {en},
	number = {4},
	urldate = {2026-03-07},
	journal = {Physical Review D},
	author = {Ferreira, Elisa G. M. and McDonough, Evan and Balkenhol, Lennart and Kallosh, Renata and Knox, Lloyd and Linde, Andrei},
	month = feb,
	year = {2026},
	pages = {043524},
}

@article{lazarides_axion_1982,
	title = {Axion models with no domain wall problem},
	volume = {115},
	copyright = {https://www.elsevier.com/tdm/userlicense/1.0/},
	issn = {03702693},
	url = {https://linkinghub.elsevier.com/retrieve/pii/0370269382905068},
	doi = {10.1016/0370-2693(82)90506-8},
	language = {en},
	number = {1},
	urldate = {2026-03-04},
	journal = {Physics Letters B},
	author = {Lazarides, G. and Shafi, Q.},
	month = aug,
	year = {1982},
	pages = {21--25},
}

@article{peccei_cp_1977,
	title = {{CP} {Conservation} in the {Presence} of {Pseudoparticles}},
	volume = {38},
	copyright = {http://link.aps.org/licenses/aps-default-license},
	issn = {0031-9007},
	url = {https://link.aps.org/doi/10.1103/PhysRevLett.38.1440},
	doi = {10.1103/PhysRevLett.38.1440},
	language = {en},
	number = {25},
	urldate = {2026-03-03},
	journal = {Physical Review Letters},
	author = {Peccei, R. D. and Quinn, Helen R.},
	month = jun,
	year = {1977},
	pages = {1440--1443},
}

@article{lee_charged_1991,
	title = {Charged black holes with scalar hair},
	volume = {44},
	copyright = {http://link.aps.org/licenses/aps-default-license},
	issn = {0556-2821},
	url = {https://link.aps.org/doi/10.1103/PhysRevD.44.3159},
	doi = {10.1103/PhysRevD.44.3159},
	language = {en},
	number = {10},
	urldate = {2026-03-03},
	journal = {Physical Review D},
	author = {Lee, Kimyeong and Weinberg, Erick J.},
	month = nov,
	year = {1991},
	pages = {3159--3163},
}

@article{gao_black_2022,
	title = {On black holes with scalar hairs},
	volume = {54},
	issn = {0001-7701, 1572-9532},
	url = {http://arxiv.org/abs/2111.11582},
	doi = {10.1007/s10714-022-03043-x},
	number = {12},
	urldate = {2026-03-02},
	journal = {General Relativity and Gravitation},
	author = {Gao, Changjun and Qiu, Jianhui},
	month = dec,
	year = {2022},
	note = {arXiv:2111.11582 [gr-qc]},
	pages = {158},
}

@misc{bizon_gravitating_1994,
	title = {Gravitating {Solitons} and {Hairy} {Black} {Holes}},
	url = {http://arxiv.org/abs/gr-qc/9402016},
	doi = {10.48550/arXiv.gr-qc/9402016},
	urldate = {2026-03-02},
	publisher = {arXiv},
	author = {Bizoń, Piotr},
	month = feb,
	year = {1994},
	note = {arXiv:gr-qc/9402016},
}

@book{misner_gravitation_2017,
	address = {Princeton, N.J},
	title = {Gravitation},
	isbn = {978-0-691-17779-3},
	publisher = {Princeton University Press},
	author = {Misner, Charles W. and Thorne, Kip S. and Wheeler, John Archibald and Kaiser, David},
	year = {2017},
}

@article{bai_primordial_2020,
	title = {Primordial {Extremal} {Black} {Holes} as {Dark} {Matter}},
	volume = {101},
	issn = {2470-0010, 2470-0029},
	url = {http://arxiv.org/abs/1906.04858},
	doi = {10.1103/PhysRevD.101.055006},
	number = {5},
	urldate = {2026-03-02},
	journal = {Physical Review D},
	author = {Bai, Yang and Orlofsky, Nicholas},
	month = mar,
	year = {2020},
	note = {arXiv:1906.04858 [hep-ph]},
	pages = {055006},
}

@article{parker_origin_1970,
	title = {The {Origin} of {Magnetic} {Fields}},
	volume = {160},
	issn = {0004-637X, 1538-4357},
	url = {http://adsabs.harvard.edu/doi/10.1086/150442},
	doi = {10.1086/150442},
	language = {en},
	urldate = {2026-02-26},
	journal = {The Astrophysical Journal},
	author = {Parker, E. N.},
	month = may,
	year = {1970},
	pages = {383},
}

@article{parker_magnetic_1987,
	title = {Magnetic monopole plasma oscillations and the survival of {Galactic} magnetic fields},
	volume = {321},
	issn = {0004-637X, 1538-4357},
	url = {http://adsabs.harvard.edu/doi/10.1086/165633},
	doi = {10.1086/165633},
	language = {en},
	urldate = {2026-02-26},
	journal = {The Astrophysical Journal},
	author = {Parker, E. N.},
	month = oct,
	year = {1987},
	pages = {349},
}

@phdthesis{perri_magnetic_2024,
	type = {{PhD} {Thesis}},
	title = {Magnetic {Monopoles} in {Cosmic} {Magnetic} {Fields}: {Acceleration} and {Constraints}},
	school = {SISSA, Trieste},
	author = {Perri, Daniele},
	year = {2024},
}

@article{bai_phenomenology_2020,
	title = {Phenomenology of {Magnetic} {Black} {Holes} with {Electroweak}-{Symmetric} {Coronas}},
	volume = {2020},
	issn = {1029-8479},
	url = {http://arxiv.org/abs/2007.03703},
	doi = {10.1007/JHEP10(2020)210},
	number = {10},
	urldate = {2026-02-26},
	journal = {Journal of High Energy Physics},
	author = {Bai, Yang and Berger, Joshua and Korwar, Mrunal and Orlofsky, Nicholas},
	month = oct,
	year = {2020},
	note = {arXiv:2007.03703 [hep-ph]},
	pages = {210},
}

@article{turner_magnetic_1982,
	title = {Magnetic monopoles and the survival of galactic magnetic fields},
	volume = {26},
	copyright = {http://link.aps.org/licenses/aps-default-license},
	issn = {0556-2821},
	url = {https://link.aps.org/doi/10.1103/PhysRevD.26.1296},
	doi = {10.1103/PhysRevD.26.1296},
	language = {en},
	number = {6},
	urldate = {2026-02-25},
	journal = {Physical Review D},
	author = {Turner, Michael S. and Parker, E. N. and Bogdan, T. J.},
	month = sep,
	year = {1982},
	pages = {1296--1305},
}

@article{burdyuzha_magnetic_2018,
	title = {Magnetic {Monopoles} and {Dark} {Matter}},
	volume = {127},
	issn = {1063-7761, 1090-6509},
	url = {http://arxiv.org/abs/1901.02341},
	doi = {10.1134/S1063776118100011},
	number = {4},
	urldate = {2026-02-16},
	journal = {Journal of Experimental and Theoretical Physics},
	author = {Burdyuzha, V. V.},
	month = oct,
	year = {2018},
	note = {arXiv:1901.02341 [physics]},
	pages = {638--646},
}

@inproceedings{department_of_physics_and_astronomy_university_of_california_irvine_california_92697_usa_magnetic_2020,
	title = {Magnetic {Monopole} {Dark} {Matter}},
	url = {http://main.andromedapublisher.com/media/img/confprocimg/2/Pdf/Final\_6503jdS.pdf},
	urldate = {2026-02-16},
	booktitle = {Proceedings of the {International} {Conference} on {Neutrinos} and {Dark} {Matter} ({NDM}-2020)},
	publisher = {Andromeda Publishing and Academic Services},
	author = {{Department of Physics and Astronomy, University of California, Irvine, California 92697, USA} and B. Verhaaren, Christopher},
	year = {2020},
}

@article{maldacena_comments_2021,
	title = {Comments on magnetic black holes},
	volume = {2021},
	issn = {1029-8479},
	url = {http://arxiv.org/abs/2004.06084},
	doi = {10.1007/JHEP04(2021)079},
	number = {4},
	urldate = {2026-02-16},
	journal = {Journal of High Energy Physics},
	author = {Maldacena, Juan},
	month = apr,
	year = {2021},
	note = {arXiv:2004.06084 [hep-th]},
	pages = {79},
}

@article{diamond_constraints_2022,
	title = {Constraints on {Relic} {Magnetic} {Black} {Holes}},
	volume = {2022},
	issn = {1029-8479},
	url = {http://arxiv.org/abs/2103.01850},
	doi = {10.1007/JHEP03(2022)157},
	number = {3},
	urldate = {2026-02-16},
	journal = {Journal of High Energy Physics},
	author = {Diamond, Melissa D. and Kaplan, David E.},
	month = mar,
	year = {2022},
	note = {arXiv:2103.01850 [hep-ph]},
	pages = {157},
}

@article{pazameta_general_2012,
	title = {A {General} {Relativistic} {Model} for {Magnetic} {Monopole}-{Infused} {Compact} {Objects}},
	volume = {339},
	issn = {0004-640X, 1572-946X},
	url = {http://arxiv.org/abs/1201.6105},
	doi = {10.1007/s10509-012-0996-7},
	number = {2},
	urldate = {2026-02-16},
	journal = {Astrophysics and Space Science},
	author = {Pazameta, Zoran},
	month = jun,
	year = {2012},
	note = {arXiv:1201.6105 [astro-ph]},
	pages = {317--322},
}

@article{gnedin_destruction_1993,
	title = {Destruction of the {Cauchy} horizon in the {Reissner}-{Nordstrom} black hole},
	volume = {10},
	issn = {0264-9381, 1361-6382},
	url = {https://iopscience.iop.org/article/10.1088/0264-9381/10/6/006},
	doi = {10.1088/0264-9381/10/6/006},
	number = {6},
	urldate = {2026-02-16},
	journal = {Classical and Quantum Gravity},
	author = {Gnedin, M L and Gnedin, N Y},
	month = jun,
	year = {1993},
	pages = {1083--1102},
}

@article{ge_sublunar-mass_2020,
	title = {Sublunar-{Mass} {Primordial} {Black} {Holes} from {Closed} {Axion} {Domain} {Walls}},
	volume = {27},
	issn = {22126864},
	url = {http://arxiv.org/abs/1905.12182},
	doi = {10.1016/j.dark.2019.100440},
	urldate = {2026-02-16},
	journal = {Physics of the Dark Universe},
	author = {Ge, Shuailiang},
	month = jan,
	year = {2020},
	note = {arXiv:1905.12182 [hep-ph]},
	pages = {100440},
}

@misc{vachaspati_lunar_2018,
	title = {Lunar {Mass} {Black} {Holes} from {QCD} {Axion} {Cosmology}},
	url = {http://arxiv.org/abs/1706.03868},
	doi = {10.48550/arXiv.1706.03868},
	urldate = {2026-02-16},
	publisher = {arXiv},
	author = {Vachaspati, Tanmay},
	month = jun,
	year = {2018},
	note = {arXiv:1706.03868 [hep-th]},
}

@article{fort_global_1993,
	title = {Do {Global} {String} {Loops} {Collapse} to {Form} {Black} {Holes}?},
	volume = {311},
	issn = {03702693},
	url = {http://arxiv.org/abs/hep-th/9305081},
	doi = {10.1016/0370-2693(93)90530-U},
	number = {1-4},
	urldate = {2026-02-16},
	journal = {Physics Letters B},
	author = {Fort, Joaquim and Vachaspati, Tanmay},
	month = jul,
	year = {1993},
	note = {arXiv:hep-th/9305081},
	pages = {41--46},
}

@article{ori_inner_1991,
	title = {Inner structure of a charged black hole: {An} exact mass-inflation solution},
	volume = {67},
	copyright = {http://link.aps.org/licenses/aps-default-license},
	issn = {0031-9007},
	shorttitle = {Inner structure of a charged black hole},
	url = {https://link.aps.org/doi/10.1103/PhysRevLett.67.789},
	doi = {10.1103/PhysRevLett.67.789},
	language = {en},
	number = {7},
	urldate = {2026-02-13},
	journal = {Physical Review Letters},
	author = {Ori, Amos},
	month = aug,
	year = {1991},
	pages = {789--792},
}

@inproceedings{cecile_m_dewitt_battelle_1968,
	title = {Battelle rencontres - 1967 lectures in mathematics and physics: {Seattle}, {WA}, {USA}, 16 - 31 {July} 1967},
	author = {{Cecile M. DeWitt} and {John A. Wheeler}},
	year = {1968},
	pages = {121--235},
}

@misc{bonanno_cauchy_2025,
	title = {Cauchy {Horizon} ({In}){Stability} of {Regular} {Black} {Holes}},
	url = {http://arxiv.org/abs/2507.03581},
	doi = {10.48550/arXiv.2507.03581},
	urldate = {2026-02-07},
	publisher = {arXiv},
	author = {Bonanno, Alfio and Panassiti, Antonio and Saueressig, Frank},
	month = jul,
	year = {2025},
	note = {arXiv:2507.03581 [gr-qc]},
}

@article{visser_dirty_1992,
	title = {Dirty blackholes: {Thermodynamics} and horizon structure},
	volume = {46},
	issn = {0556-2821},
	shorttitle = {Dirty blackholes},
	url = {http://arxiv.org/abs/hep-th/9203057},
	doi = {10.1103/PhysRevD.46.2445},
	number = {6},
	urldate = {2026-02-07},
	journal = {Physical Review D},
	author = {Visser, Matt},
	month = sep,
	year = {1992},
	note = {arXiv:hep-th/9203057},
	pages = {2445--2451},
}

@article{coleman_quantum_1992,
	title = {Quantum {Hair} on {Black} {Holes}},
	volume = {378},
	issn = {05503213},
	url = {http://arxiv.org/abs/hep-th/9201059},
	doi = {10.1016/0550-3213(92)90008-Y},
	number = {1-2},
	urldate = {2026-02-07},
	journal = {Nuclear Physics B},
	author = {Coleman, Sidney and Preskill, John and Wilczek, Frank},
	month = jul,
	year = {1992},
	note = {arXiv:hep-th/9201059},
	pages = {175--246},
}

@article{maeda_novel_2005,
	title = {Novel {Cauchy}-horizon instability},
	volume = {71},
	copyright = {http://link.aps.org/licenses/aps-default-license},
	issn = {1550-7998, 1550-2368},
	url = {https://link.aps.org/doi/10.1103/PhysRevD.71.064015},
	doi = {10.1103/PhysRevD.71.064015},
	language = {en},
	number = {6},
	urldate = {2026-02-07},
	journal = {Physical Review D},
	author = {Maeda, Hideki and Torii, Takashi and Harada, Tomohiro},
	month = mar,
	year = {2005},
	pages = {064015},
}

@article{poisson_internal_1990,
	title = {Internal structure of black holes},
	volume = {41},
	copyright = {http://link.aps.org/licenses/aps-default-license},
	issn = {0556-2821},
	url = {https://link.aps.org/doi/10.1103/PhysRevD.41.1796},
	doi = {10.1103/PhysRevD.41.1796},
	language = {en},
	number = {6},
	urldate = {2026-02-07},
	journal = {Physical Review D},
	author = {Poisson, Eric and Israel, Werner},
	month = mar,
	year = {1990},
	pages = {1796--1809},
}

@article{bai_hairy_2021,
	title = {Hairy magnetic and dyonic black holes in the {Standard} {Model}},
	volume = {2021},
	issn = {1029-8479},
	url = {https://link.springer.com/10.1007/JHEP04(2021)119},
	doi = {10.1007/JHEP04(2021)119},
	language = {en},
	number = {4},
	urldate = {2026-02-07},
	journal = {Journal of High Energy Physics},
	author = {Bai, Yang and Korwar, Mrunal},
	month = apr,
	year = {2021},
	pages = {119},
}

@article{barriola_gravitational_1989,
	title = {Gravitational field of a global monopole},
	volume = {63},
	copyright = {http://link.aps.org/licenses/aps-default-license},
	issn = {0031-9007},
	url = {https://link.aps.org/doi/10.1103/PhysRevLett.63.341},
	doi = {10.1103/PhysRevLett.63.341},
	language = {en},
	number = {4},
	urldate = {2026-02-05},
	journal = {Physical Review Letters},
	author = {Barriola, Manuel and Vilenkin, Alexander},
	month = jul,
	year = {1989},
	pages = {341--343},
}

@article{vilenkin_topological_1994,
	title = {Topological inflation},
	volume = {72},
	copyright = {http://link.aps.org/licenses/aps-default-license},
	issn = {0031-9007},
	url = {https://link.aps.org/doi/10.1103/PhysRevLett.72.3137},
	doi = {10.1103/PhysRevLett.72.3137},
	language = {en},
	number = {20},
	urldate = {2026-02-05},
	journal = {Physical Review Letters},
	author = {Vilenkin, Alexander},
	month = may,
	year = {1994},
	pages = {3137--3140},
}

@article{polyakov_particle_1974,
	title = {Particle spectrum in quantum field theory},
	volume = {20},
	issn = {0021-3640},
	url = {https://ui.adsabs.harvard.edu/abs/1974JETPL..20..194P},
	urldate = {2026-02-04},
	journal = {Soviet Journal of Experimental and Theoretical Physics Letters},
	publisher = {Springer},
	author = {Polyakov, A. M.},
	month = sep,
	year = {1974},
	note = {ADS Bibcode: 1974JETPL..20..194P},
	pages = {194},
}

@article{hooft_magnetic_1974,
	title = {Magnetic monopoles in unified gauge theories},
	volume = {79},
	copyright = {https://www.elsevier.com/tdm/userlicense/1.0/},
	issn = {05503213},
	url = {https://linkinghub.elsevier.com/retrieve/pii/0550321374904866},
	doi = {10.1016/0550-3213(74)90486-6},
	language = {en},
	number = {2},
	urldate = {2026-02-04},
	journal = {Nuclear Physics B},
	author = {Hooft, G.'t},
	month = sep,
	year = {1974},
	pages = {276--284},
}

@article{rajantie_magnetic_2003,
	title = {Magnetic monopoles from gauge theory phase transitions},
	volume = {68},
	issn = {0556-2821, 1089-4918},
	url = {http://arxiv.org/abs/hep-ph/0212130},
	doi = {10.1103/PhysRevD.68.021301},
	number = {2},
	urldate = {2026-02-04},
	journal = {Physical Review D},
	author = {Rajantie, A.},
	month = jul,
	year = {2003},
	note = {arXiv:hep-ph/0212130},
	pages = {021301},
}

@article{kasuya_topological_1998,
	title = {Topological {Defects} {Formation} after {Inflation} on {Lattice} {Simulation}},
	volume = {58},
	issn = {0556-2821, 1089-4918},
	url = {http://arxiv.org/abs/hep-ph/9804429},
	doi = {10.1103/PhysRevD.58.083516},
	number = {8},
	urldate = {2026-02-04},
	journal = {Physical Review D},
	author = {Kasuya, S. and Kawasaki, M.},
	month = sep,
	year = {1998},
	note = {arXiv:hep-ph/9804429},
	pages = {083516},
}

@article{bolognesi_magnetic_2011,
	title = {Magnetic {Bags} and {Black} {Holes}},
	volume = {845},
	issn = {05503213},
	url = {http://arxiv.org/abs/1005.4642},
	doi = {10.1016/j.nuclphysb.2010.12.008},
	number = {3},
	urldate = {2026-01-29},
	journal = {Nuclear Physics B},
	author = {Bolognesi, S.},
	month = apr,
	year = {2011},
	note = {arXiv:1005.4642 [hep-th]},
	pages = {324--339},
}

@article{weinberg_classical_1992,
	title = {Classical {Solutions} in {Quantum} {Field} {Theories}},
	volume = {42},
	issn = {0163-8998, 1545-4134},
	url = {https://www.annualreviews.org/doi/10.1146/annurev.ns.42.120192.001141},
	doi = {10.1146/annurev.ns.42.120192.001141},
	language = {en},
	number = {1},
	urldate = {2026-01-29},
	journal = {Annual Review of Nuclear and Particle Science},
	author = {Weinberg, E J},
	month = dec,
	year = {1992},
	pages = {177--210},
}

@article{tong_tasi_nodate,
	title = {{TASI} {Lectures} on {Solitons}},
	language = {en},
	author = {Tong, David},
}

@article{mazur_gravitational_2004,
	title = {Gravitational {Vacuum} {Condensate} {Stars}},
	volume = {101},
	issn = {0027-8424, 1091-6490},
	url = {http://arxiv.org/abs/gr-qc/0407075},
	doi = {10.1073/pnas.0402717101},
	number = {26},
	urldate = {2026-01-17},
	journal = {Proceedings of the National Academy of Sciences},
	author = {Mazur, Pawel O. and Mottola, Emil},
	month = jun,
	year = {2004},
	note = {arXiv:gr-qc/0407075},
	pages = {9545--9550},
}

@article{breitenlohner_gravitating_1992,
	title = {Gravitating monopole solutions},
	volume = {383},
	copyright = {https://www.elsevier.com/tdm/userlicense/1.0/},
	issn = {05503213},
	url = {https://linkinghub.elsevier.com/retrieve/pii/0550321392906822},
	doi = {10.1016/0550-3213(92)90682-2},
	language = {en},
	number = {1-2},
	urldate = {2025-12-24},
	journal = {Nuclear Physics B},
	author = {Breitenlohner, Peter and Forgács, Peter and Maison, Dieter},
	month = sep,
	year = {1992},
	pages = {357--376},
}

@misc{bokulic_conundrum_2025,
	title = {Conundrum of regular black holes with nonlinear electromagnetic fields},
	url = {http://arxiv.org/abs/2510.23711},
	doi = {10.48550/arXiv.2510.23711},
	urldate = {2025-12-19},
	publisher = {arXiv},
	author = {Bokulić, Ana and Jurić, Tajron and Smolić, Ivica},
	month = oct,
	year = {2025},
	note = {arXiv:2510.23711 [gr-qc]},
}

@article{arreaga_stability_2000,
	title = {Stability of self-gravitating magnetic monopoles},
	volume = {62},
	issn = {0556-2821, 1089-4918},
	url = {http://arxiv.org/abs/gr-qc/0001078},
	doi = {10.1103/PhysRevD.62.043520},
	number = {4},
	urldate = {2025-12-12},
	journal = {Physical Review D},
	author = {Arreaga, Guillermo and Cho, Inyong and Guven, Jemal},
	month = jul,
	year = {2000},
	note = {arXiv:gr-qc/0001078},
	pages = {043520},
}

@misc{felice_exotic_2025,
	title = {Exotic compact objects in {Einstein}-{Scalar}-{Maxwell} theories},
	url = {http://arxiv.org/abs/2511.14207},
	doi = {10.48550/arXiv.2511.14207},
	urldate = {2025-12-12},
	publisher = {arXiv},
	author = {Felice, Antonio De and Tsujikawa, Shinji},
	month = nov,
	year = {2025},
	note = {arXiv:2511.14207 [gr-qc]},
}

@article{ramirez-valdez_dyonic_2023,
	title = {Dyonic black holes: {The} theory of two electromagnetic potentials},
	volume = {107},
	issn = {2470-0010, 2470-0029},
	shorttitle = {Dyonic black holes},
	url = {http://arxiv.org/abs/2301.09330},
	doi = {10.1103/PhysRevD.107.064016},
	number = {6},
	urldate = {2025-12-12},
	journal = {Physical Review D},
	author = {Ramírez-Valdez, C. J. and García-Compeán, H. and Manko, V. S.},
	month = mar,
	year = {2023},
	note = {arXiv:2301.09330 [gr-qc]},
	pages = {064016},
}

@article{felice_instability_2025,
	title = {Instability of nonsingular black holes in nonlinear electrodynamics},
	volume = {134},
	issn = {0031-9007, 1079-7114},
	url = {http://arxiv.org/abs/2410.00314},
	doi = {10.1103/PhysRevLett.134.081401},
	number = {8},
	urldate = {2025-12-12},
	journal = {Physical Review Letters},
	author = {Felice, Antonio De and Tsujikawa, Shinji},
	month = feb,
	year = {2025},
	note = {arXiv:2410.00314 [gr-qc]},
	pages = {081401},
}

@article{kleihaus_stationary_2008,
	title = {Stationary {Dyonic} {Regular} and {Black} {Hole} {Solutions}},
	volume = {40},
	issn = {0001-7701, 1572-9532},
	url = {http://arxiv.org/abs/0705.1511},
	doi = {10.1007/s10714-007-0604-2},
	number = {6},
	urldate = {2025-12-12},
	journal = {General Relativity and Gravitation},
	author = {Kleihaus, Burkhard and Kunz, Jutta and Navarro-Lérida, Francisco and Neemann, Ulrike},
	month = jun,
	year = {2008},
	note = {arXiv:0705.1511 [gr-qc]},
	pages = {1279--1310},
}

@article{bronnikov_regular_2001,
	title = {Regular {Magnetic} {Black} {Holes} and {Monopoles} from {Nonlinear} {Electrodynamics}},
	volume = {63},
	issn = {0556-2821, 1089-4918},
	url = {http://arxiv.org/abs/gr-qc/0006014},
	doi = {10.1103/PhysRevD.63.044005},
	number = {4},
	urldate = {2025-12-12},
	journal = {Physical Review D},
	author = {Bronnikov, Kirill A.},
	month = jan,
	year = {2001},
	note = {arXiv:gr-qc/0006014},
	pages = {044005},
}

@article{junior_dyonic_2025,
	title = {Dyonic regular black bounce solutions in {General} {Relativity}},
	volume = {85},
	issn = {1434-6052},
	url = {http://arxiv.org/abs/2502.13327},
	doi = {10.1140/epjc/s10052-025-14427-z},
	number = {7},
	urldate = {2025-12-12},
	journal = {The European Physical Journal C},
	author = {Junior, Ednaldo L. B. and Junior, José Tarciso S. S. and Lobo, Francisco S. N. and Rodrigues, Manuel E. and Silva, Luís F. Dias da and Vieira, Henrique A.},
	month = jul,
	year = {2025},
	note = {arXiv:2502.13327 [gr-qc]},
	pages = {724},
}

@article{sueto_evaporation_2023,
	title = {Evaporation of a nonsingular {Reissner}-{Nordström} black hole and information loss problem},
	volume = {2023},
	issn = {2050-3911},
	url = {http://arxiv.org/abs/2301.10456},
	doi = {10.1093/ptep/ptad111},
	number = {10},
	urldate = {2025-12-12},
	journal = {Progress of Theoretical and Experimental Physics},
	author = {Sueto, Kensuke and Yoshino, Hirotaka},
	month = oct,
	year = {2023},
	note = {arXiv:2301.10456 [gr-qc]},
	pages = {103E01},
}

@article{hayward_formation_2006,
	title = {Formation and evaporation of non-singular black holes},
	volume = {96},
	issn = {0031-9007, 1079-7114},
	url = {http://arxiv.org/abs/gr-qc/0506126},
	doi = {10.1103/PhysRevLett.96.031103},
	number = {3},
	urldate = {2025-12-12},
	journal = {Physical Review Letters},
	author = {Hayward, Sean A.},
	month = jan,
	year = {2006},
	note = {arXiv:gr-qc/0506126},
	pages = {031103},
}

@article{mbonye_nonsingular_2005,
	title = {Nonsingular black hole model as a possible end product of gravitational collapse},
	volume = {72},
	copyright = {http://link.aps.org/licenses/aps-default-license},
	issn = {1550-7998, 1550-2368},
	url = {https://link.aps.org/doi/10.1103/PhysRevD.72.024016},
	doi = {10.1103/PhysRevD.72.024016},
	language = {en},
	number = {2},
	urldate = {2025-12-12},
	journal = {Physical Review D},
	author = {Mbonye, Manasse R. and Kazanas, Demosthenes},
	month = jul,
	year = {2005},
	pages = {024016},
}

@article{davies_nonsingular_2025,
	title = {Nonsingular black holes as dark matter},
	volume = {111},
	issn = {2470-0010, 2470-0029},
	url = {https://link.aps.org/doi/10.1103/PhysRevD.111.103512},
	doi = {10.1103/PhysRevD.111.103512},
	language = {en},
	number = {10},
	urldate = {2025-12-12},
	journal = {Physical Review D},
	author = {Davies, Paul C. W. and Easson, Damien A. and Levin, Phillip B.},
	month = may,
	year = {2025},
	pages = {103512},
}

@article{ayon-beato_bardeen_2000,
	title = {The {Bardeen} {Model} as a {Nonlinear} {Magnetic} {Monopole}},
	volume = {493},
	issn = {03702693},
	url = {http://arxiv.org/abs/gr-qc/0009077},
	doi = {10.1016/S0370-2693(00)01125-4},
	number = {1-2},
	urldate = {2025-12-12},
	journal = {Physics Letters B},
	author = {Ayón-Beato, Eloy and García, Alberto},
	month = nov,
	year = {2000},
	note = {arXiv:gr-qc/0009077},
	pages = {149--152},
}

@book{cheng_gauge_2011,
	address = {Oxford},
	edition = {Reprinted with corrections},
	series = {Oxford science publications},
	title = {Gauge theory of elementary particle physics},
	isbn = {978-0-19-851961-4},
	language = {eng},
	publisher = {Clarendon Press},
	author = {Cheng, Ta-Pei and Li, Ling-Fong},
	year = {2011},
}

@article{preskill_magnetic_1984,
	title = {Magnetic {Monopoles}},
	volume = {34},
	issn = {0163-8998, 1545-4134},
	url = {https://www.annualreviews.org/doi/10.1146/annurev.ns.34.120184.002333},
	doi = {10.1146/annurev.ns.34.120184.002333},
	language = {en},
	number = {1},
	urldate = {2025-11-11},
	journal = {Annual Review of Nuclear and Particle Science},
	author = {Preskill, J},
	month = dec,
	year = {1984},
	pages = {461--530},
}

@article{schwinger_magnetic_1966,
	title = {Magnetic {Charge} and {Quantum} {Field} {Theory}},
	volume = {144},
	copyright = {http://link.aps.org/licenses/aps-default-license},
	issn = {0031-899X},
	url = {https://link.aps.org/doi/10.1103/PhysRev.144.1087},
	doi = {10.1103/PhysRev.144.1087},
	language = {en},
	number = {4},
	urldate = {2025-11-04},
	journal = {Physical Review},
	author = {Schwinger, Julian},
	month = apr,
	year = {1966},
	pages = {1087--1093},
}

@article{zwanziger_quantum_1968,
	title = {Quantum {Field} {Theory} of {Particles} with {Both} {Electric} and {Magnetic} {Charges}},
	volume = {176},
	copyright = {http://link.aps.org/licenses/aps-default-license},
	issn = {0031-899X},
	url = {https://link.aps.org/doi/10.1103/PhysRev.176.1489},
	doi = {10.1103/PhysRev.176.1489},
	language = {en},
	number = {5},
	urldate = {2025-11-04},
	journal = {Physical Review},
	author = {Zwanziger, Daniel},
	month = dec,
	year = {1968},
	pages = {1489--1495},
}

@article{dirac_quantised_1931,
	title = {Quantised singularities in the electromagnetic field},
	volume = {133},
	copyright = {https://royalsociety.org/journals/ethics-policies/data-sharing-mining/},
	issn = {0950-1207, 2053-9150},
	url = {https://royalsocietypublishing.org/doi/10.1098/rspa.1931.0130},
	doi = {10.1098/rspa.1931.0130},
	language = {en},
	number = {821},
	urldate = {2025-11-04},
	journal = {Proceedings of the Royal Society of London. Series A, Containing Papers of a Mathematical and Physical Character},
	author = {Dirac, Paul A. M.},
	month = sep,
	year = {1931},
	pages = {60--72},
}

@article{sato_unified_2018,
	title = {Unified origin of axion and monopole dark matter, and solution to the domain-wall problem},
	volume = {98},
	issn = {2470-0010, 2470-0029},
	url = {https://link.aps.org/doi/10.1103/PhysRevD.98.043535},
	doi = {10.1103/PhysRevD.98.043535},
	language = {en},
	number = {4},
	urldate = {2025-10-24},
	journal = {Physical Review D},
	author = {Sato, Ryosuke and Takahashi, Fuminobu and Yamada, Masaki},
	month = aug,
	year = {2018},
	pages = {043535},
}

@article{wilczek_two_1987,
	title = {Two applications of axion electrodynamics},
	volume = {58},
	copyright = {http://link.aps.org/licenses/aps-default-license},
	issn = {0031-9007},
	url = {https://link.aps.org/doi/10.1103/PhysRevLett.58.1799},
	doi = {10.1103/PhysRevLett.58.1799},
	language = {en},
	number = {18},
	urldate = {2025-10-14},
	journal = {Physical Review Letters},
	author = {Wilczek, Frank},
	month = may,
	year = {1987},
	pages = {1799--1802},
}

@misc{kogan_axions_1993,
	title = {Axions, {Monopoles} and {Cosmic} {String}},
	url = {http://arxiv.org/abs/hep-ph/9305307},
	doi = {10.48550/arXiv.hep-ph/9305307},
	urldate = {2025-10-14},
	publisher = {arXiv},
	author = {Kogan, Ian I.},
	month = may,
	year = {1993},
	note = {arXiv:hep-ph/9305307},
}

@article{lee_topological_1987,
	title = {Topological mass terms on axion domain walls},
	volume = {35},
	copyright = {http://link.aps.org/licenses/aps-default-license},
	issn = {0556-2821},
	url = {https://link.aps.org/doi/10.1103/PhysRevD.35.3286},
	doi = {10.1103/PhysRevD.35.3286},
	language = {en},
	number = {10},
	urldate = {2025-10-14},
	journal = {Physical Review D},
	author = {Lee, Kimyeong},
	month = may,
	year = {1987},
	pages = {3286--3289},
}

@article{kogan_kaluza-klein_1993,
	title = {Kaluza-{Klein} and axion domain walls. {Induced} charge and mass transmutation},
	volume = {299},
	copyright = {https://www.elsevier.com/tdm/userlicense/1.0/},
	issn = {03702693},
	url = {https://linkinghub.elsevier.com/retrieve/pii/037026939390877K},
	doi = {10.1016/0370-2693(93)90877-K},
	language = {en},
	number = {1-2},
	urldate = {2025-10-13},
	journal = {Physics Letters B},
	author = {Kogan, Ian I.},
	month = jan,
	year = {1993},
	pages = {16--23},
}

@article{witten_dyons_1979,
	title = {Dyons of charge e$\theta$/2$\pi$},
	volume = {86},
	copyright = {https://www.elsevier.com/tdm/userlicense/1.0/},
	issn = {03702693},
	url = {https://linkinghub.elsevier.com/retrieve/pii/0370269379908384},
	doi = {10.1016/0370-2693(79)90838-4},
	language = {en},
	number = {3-4},
	urldate = {2025-10-13},
	journal = {Physics Letters B},
	author = {Witten, E.},
	month = oct,
	year = {1979},
	pages = {283--287},
}

@misc{dine_remarks_2023,
	title = {Remarks on the {Axion} {Domain} {Wall} {Problem}},
	url = {http://arxiv.org/abs/2307.04710},
	doi = {10.48550/arXiv.2307.04710},
	urldate = {2025-10-07},
	publisher = {arXiv},
	author = {Dine, Michael},
	month = jul,
	year = {2023},
	note = {arXiv:2307.04710 [hep-ph]},
}

@article{vilenkin_cosmic_1982,
	title = {Cosmic {Strings} and {Domain} {Walls} in {Models} with {Goldstone} and {Pseudo}-{Goldstone} {Bosons}},
	volume = {48},
	copyright = {http://link.aps.org/licenses/aps-default-license},
	issn = {0031-9007},
	url = {https://link.aps.org/doi/10.1103/PhysRevLett.48.1867},
	doi = {10.1103/PhysRevLett.48.1867},
	language = {en},
	number = {26},
	urldate = {2025-10-01},
	journal = {Physical Review Letters},
	author = {Vilenkin, Alexander and Everett, Allen E.},
	month = jun,
	year = {1982},
	pages = {1867--1870},
}

@article{dunsky_primordial_2024,
	title = {Primordial {Black} {Holes} from {Axion} {Domain} {Wall} {Collapse}},
	volume = {2024},
	issn = {1029-8479},
	url = {http://arxiv.org/abs/2402.03426},
	doi = {10.1007/JHEP06(2024)198},
	number = {6},
	urldate = {2025-10-01},
	journal = {Journal of High Energy Physics},
	author = {Dunsky, David I. and Kongsore, Marius},
	month = jun,
	year = {2024},
	note = {arXiv:2402.03426 [hep-ph]},
	pages = {198},
}

@incollection{peccei_strong_2008,
	title = {The {Strong} {CP} {Problem} and {Axions}},
	volume = {741},
	url = {http://arxiv.org/abs/hep-ph/0607268},
	urldate = {2025-10-01},
	author = {Peccei, R. D.},
	year = {2008},
	note = {arXiv:hep-ph/0607268},
	pages = {3--17},
}

@article{fischler_dyon-axion_1983,
	title = {Dyon-axion dynamics},
	volume = {125},
	copyright = {https://www.elsevier.com/tdm/userlicense/1.0/},
	issn = {03702693},
	url = {https://linkinghub.elsevier.com/retrieve/pii/0370269383912601},
	doi = {10.1016/0370-2693(83)91260-1},
	language = {en},
	number = {2-3},
	urldate = {2025-09-29},
	journal = {Physics Letters B},
	author = {Fischler, Willy and Preskill, John},
	month = may,
	year = {1983},
	pages = {165--170},
}

@article{sikivie_interaction_1984,
	title = {On the interaction of magnetic monopoles with axionic domain walls},
	volume = {137},
	copyright = {https://www.elsevier.com/tdm/userlicense/1.0/},
	issn = {03702693},
	url = {https://linkinghub.elsevier.com/retrieve/pii/0370269384917313},
	doi = {10.1016/0370-2693(84)91731-3},
	language = {en},
	number = {5-6},
	urldate = {2025-09-29},
	journal = {Physics Letters B},
	author = {Sikivie, P.},
	month = apr,
	year = {1984},
	pages = {353--356},
}

@article{beltracchi_slowly_2022,
	title = {Slowly rotating gravastars},
	volume = {105},
	issn = {2470-0010, 2470-0029},
	url = {http://arxiv.org/abs/2107.00762},
	doi = {10.1103/PhysRevD.105.024002},
	number = {2},
	urldate = {2025-09-28},
	journal = {Physical Review D},
	author = {Beltracchi, Philip and Gondolo, Paolo and Mottola, Emil},
	month = jan,
	year = {2022},
	note = {arXiv:2107.00762 [gr-qc]},
	pages = {024002},
}

@article{uchikata_slowly_2016,
	title = {Slowly rotating thin shell gravastars},
	volume = {33},
	issn = {0264-9381, 1361-6382},
	url = {http://arxiv.org/abs/1506.06485},
	doi = {10.1088/0264-9381/33/2/025005},
	number = {2},
	urldate = {2025-09-28},
	journal = {Classical and Quantum Gravity},
	author = {Uchikata, Nami and Yoshida, Shijun},
	month = jan,
	year = {2016},
	note = {arXiv:1506.06485 [gr-qc]},
	pages = {025005},
}

@article{huang_structure_1985,
	title = {Structure of axionic domain walls},
	volume = {32},
	copyright = {http://link.aps.org/licenses/aps-default-license},
	issn = {0556-2821},
	url = {https://link.aps.org/doi/10.1103/PhysRevD.32.1560},
	doi = {10.1103/PhysRevD.32.1560},
	language = {en},
	number = {6},
	urldate = {2025-09-27},
	journal = {Physical Review D},
	author = {Huang, M. C. and Sikivie, P.},
	month = sep,
	year = {1985},
	pages = {1560--1568},
}

@article{vilenkin_cosmic_1985,
	title = {Cosmic strings and domain walls},
	volume = {121},
	copyright = {https://www.elsevier.com/tdm/userlicense/1.0/},
	issn = {0370-1573},
	url = {https://linkinghub.elsevier.com/retrieve/pii/037015738590033X},
	doi = {10.1016/0370-1573(85)90033-x},
	language = {en},
	number = {5},
	urldate = {2025-09-03},
	journal = {Physics Reports},
	publisher = {Elsevier BV},
	author = {Vilenkin, Alexander},
	month = may,
	year = {1985},
	pages = {263--315},
}

@article{silveira_dynamics_1988,
	title = {Dynamics of the $\lambda$$\phi$4kink},
	volume = {38},
	copyright = {http://link.aps.org/licenses/aps-default-license},
	issn = {0556-2821},
	url = {https://link.aps.org/doi/10.1103/PhysRevD.38.3823},
	doi = {10.1103/physrevd.38.3823},
	language = {en},
	number = {12},
	urldate = {2025-08-27},
	journal = {Physical Review D},
	publisher = {American Physical Society (APS)},
	author = {Silveira, Vanda},
	month = dec,
	year = {1988},
	pages = {3823--3826},
}

@article{widrow_dynamics_1989,
	title = {Dynamics of thick domain walls},
	volume = {40},
	copyright = {http://link.aps.org/licenses/aps-default-license},
	issn = {0556-2821},
	url = {https://link.aps.org/doi/10.1103/PhysRevD.40.1002},
	doi = {10.1103/physrevd.40.1002},
	language = {en},
	number = {4},
	urldate = {2025-08-27},
	journal = {Physical Review D},
	publisher = {American Physical Society (APS)},
	author = {Widrow, Lawrence M.},
	month = aug,
	year = {1989},
	pages = {1002--1010},
}

@article{hiramatsu_production_2012,
	title = {Production of dark matter axions from collapse of string-wall systems},
	volume = {85},
	copyright = {http://link.aps.org/licenses/aps-default-license},
	issn = {1550-7998, 1550-2368},
	url = {https://link.aps.org/doi/10.1103/PhysRevD.85.105020},
	doi = {10.1103/physrevd.85.105020},
	language = {en},
	number = {10},
	urldate = {2025-08-27},
	journal = {Physical Review D},
	publisher = {American Physical Society (APS)},
	author = {Hiramatsu, Takashi and Kawasaki, Masahiro and Saikawa, Ken’ichi and Sekiguchi, Toyokazu},
	month = may,
	year = {2012},
}

@article{hiramatsu_evolution_2011,
	title = {Evolution of {String}-{Wall} {Networks} and {Axionic} {Domain} {Wall} {Problem}},
	volume = {2011},
	issn = {1475-7516},
	url = {http://arxiv.org/abs/1012.4558},
	doi = {10.1088/1475-7516/2011/08/030},
	number = {08},
	urldate = {2025-08-27},
	journal = {Journal of Cosmology and Astroparticle Physics},
	author = {Hiramatsu, Takashi and Kawasaki, Masahiro and Saikawa, Ken'ichi},
	month = aug,
	year = {2011},
	note = {arXiv:1012.4558 [astro-ph]},
	pages = {030--030},
}

@article{hiramatsu_axion_2013,
	title = {Axion cosmology with long-lived domain walls},
	volume = {2013},
	issn = {1475-7516},
	url = {http://arxiv.org/abs/1207.3166},
	doi = {10.1088/1475-7516/2013/01/001},
	number = {01},
	urldate = {2025-08-22},
	journal = {Journal of Cosmology and Astroparticle Physics},
	author = {Hiramatsu, Takashi and Kawasaki, Masahiro and Saikawa, Ken'ichi and Sekiguchi, Toyokazu},
	month = jan,
	year = {2013},
	note = {arXiv:1207.3166 [hep-ph]},
	pages = {001--001},
}

@article{gelmini_primordial_2023,
	title = {Primordial black hole dark matter from catastrogenesis with unstable pseudo-{Goldstone} bosons},
	volume = {2023},
	issn = {1475-7516},
	url = {http://arxiv.org/abs/2303.14107},
	doi = {10.1088/1475-7516/2023/06/055},
	number = {06},
	urldate = {2025-08-22},
	journal = {Journal of Cosmology and Astroparticle Physics},
	author = {Gelmini, Graciela B. and Hyman, Jonah and Simpson, Anna and Vitagliano, Edoardo},
	month = jun,
	year = {2023},
	note = {arXiv:2303.14107 [hep-ph]},
	pages = {055},
}

@article{tanahashi_spherical_2015,
	title = {Spherical {Domain} {Wall} {Collapse} in a {Dust} {Universe}},
	volume = {32},
	issn = {0264-9381, 1361-6382},
	url = {http://arxiv.org/abs/1411.7479},
	doi = {10.1088/0264-9381/32/15/155003},
	number = {15},
	urldate = {2025-07-29},
	journal = {Classical and Quantum Gravity},
	author = {Tanahashi, Norihiro and Yoo, Chul-Moon},
	month = aug,
	year = {2015},
	note = {arXiv:1411.7479 [gr-qc]},
	pages = {155003},
}

@article{widrow_collapse_1989,
	title = {Collapse of nearly spherical domain walls},
	volume = {39},
	copyright = {http://link.aps.org/licenses/aps-default-license},
	issn = {0556-2821},
	url = {https://link.aps.org/doi/10.1103/PhysRevD.39.3576},
	doi = {10.1103/physrevd.39.3576},
	language = {en},
	number = {12},
	urldate = {2025-07-30},
	journal = {Physical Review D},
	publisher = {American Physical Society (APS)},
	author = {Widrow, Lawrence M.},
	month = jun,
	year = {1989},
	pages = {3576--3578},
}

@book{vilenkin_cosmic_2001,
	address = {Cambridge [u.a.]: Cambridge Univ. Press},
	edition = {Reprint},
	series = {Cambridge monographs on mathematical physics},
	title = {Cosmic strings and other topological defects},
	isbn = {978-0-521-65476-0 978-0-521-39153-5},
	language = {eng},
	author = {Vilenkin, Alex and Shellard, E. P. S.},
	year = {2001},
}

@article{gross_magnetic_1983,
    title = {Magnetic monopoles in {Kaluza}-{Klein} theories},
    volume = {226},
    copyright = {https://www.elsevier.com/tdm/userlicense/1.0/},
    issn = {05503213},
    url = {https://linkinghub.elsevier.com/retrieve/pii/0550321383904625},
    doi = {10.1016/0550-3213(83)90462-5},
    language = {en},
    number = {1},
    urldate = {2026-06-25},
    journal = {Nuclear Physics B},
    author = {Gross, David J. and Perry, Malcolm J.},
    month = sep,
    year = {1983},
    pages = {29--48},
}

@article{sorkin_kaluza-klein_1983,
    title = {Kaluza-{Klein} {Monopole}},
    volume = {51},
    copyright = {http://link.aps.org/licenses/aps-default-license},
    issn = {0031-9007},
    url = {https://link.aps.org/doi/10.1103/PhysRevLett.51.87},
    doi = {10.1103/PhysRevLett.51.87},
    language = {en},
    number = {2},
    urldate = {2026-06-25},
    journal = {Physical Review Letters},
    author = {Sorkin, Rafael D.},
    month = jul,
    year = {1983},
    pages = {87--90},
}

@article{diaconescu_d-branes_1997,
    title = {D-branes, {Monopoles} and {Nahm} {Equations}},
    volume = {503},
    issn = {05503213},
    url = {http://arxiv.org/abs/hep-th/9608163},
    doi = {10.1016/S0550-3213(97)00438-0},
    number = {1-2},
    urldate = {2026-06-25},
    journal = {Nuclear Physics B},
    author = {Diaconescu, Duiliu-Emanuel},
    month = oct,
    year = {1997},
    note = {arXiv:hep-th/9608163},
    pages = {220--238},
}

@misc{verlinde_metastable_2006,
    title = {On {Metastable} {Branes} and a {New} {Type} of {Magnetic} {Monopole}},
    url = {http://arxiv.org/abs/hep-th/0611069},
    doi = {10.48550/arXiv.hep-th/0611069},
    urldate = {2026-06-25},
    publisher = {arXiv},
    author = {Verlinde, Herman},
    month = nov,
    year = {2006},
    note = {arXiv:hep-th/0611069},
}

@incollection{sikivie_axion_2008,
    title = {Axion {Cosmology}},
    volume = {741},
    url = {http://arxiv.org/abs/astro-ph/0610440},
    urldate = {2026-06-26},
    author = {Sikivie, Pierre},
    year = {2008},
    note = {arXiv:astro-ph/0610440},
    pages = {19--50},
}

@article{arvanitaki_string_2010,
    title = {String {Axiverse}},
    volume = {81},
    issn = {1550-7998, 1550-2368},
    url = {http://arxiv.org/abs/0905.4720},
    doi = {10.1103/PhysRevD.81.123530},
    number = {12},
    urldate = {2026-07-03},
    journal = {Physical Review D},
    author = {Arvanitaki, Asimina and Dimopoulos, Savas and Dubovsky, Sergei and Kaloper, Nemanja and March-Russell, John},
    month = jun,
    year = {2010},
    note = {arXiv:0905.4720 [hep-th]},
    pages = {123530},
}

@article{conlon_qcd_2006,
    title = {The {QCD} {Axion} and {Moduli} {Stabilisation}},
    volume = {2006},
    issn = {1029-8479},
    url = {http://arxiv.org/abs/hep-th/0602233},
    doi = {10.1088/1126-6708/2006/05/078},
    number = {05},
    urldate = {2026-07-03},
    journal = {Journal of High Energy Physics},
    author = {Conlon, Joseph P.},
    month = jun,
    year = {2006},
    note = {arXiv:hep-th/0602233},
    pages = {078--078},
}

@article{cortona_qcd_2016,
    title = {The {QCD} axion, precisely},
    volume = {2016},
    issn = {1029-8479},
    url = {http://arxiv.org/abs/1511.02867},
    doi = {10.1007/JHEP01(2016)034},
    number = {1},
    urldate = {2026-07-03},
    journal = {Journal of High Energy Physics},
    author = {Cortona, Giovanni Grilli di and Hardy, Edward and Vega, Javier Pardo and Villadoro, Giovanni},
    month = jan,
    year = {2016},
    note = {arXiv:1511.02867 [hep-ph]},
    pages = {34},
}

@article{svrcek_axions_2006,
    title = {Axions {In} {String} {Theory}},
    volume = {2006},
    issn = {1029-8479},
    url = {http://arxiv.org/abs/hep-th/0605206},
    doi = {10.1088/1126-6708/2006/06/051},
    number = {06},
    urldate = {2026-07-03},
    journal = {Journal of High Energy Physics},
    author = {Svrcek, Peter and Witten, Edward},
    month = jun,
    year = {2006},
    note = {arXiv:hep-th/0605206},
    pages = {051--051},
}

@article{kim_light_1987,
    title = {Light pseudoscalars, particle physics and cosmology},
    volume = {150},
    issn = {03701573},
    url = {https://linkinghub.elsevier.com/retrieve/pii/0370157387900172},
    doi = {10.1016/0370-1573(87)90017-2},
    language = {en},
    number = {1-2},
    urldate = {2026-07-03},
    journal = {Physics Reports},
    author = {Kim, Jihn E.},
    month = jun,
    year = {1987},
    pages = {1--177},
}

@misc{borsanyi_lattice_2016,
    title = {Lattice {QCD} for {Cosmology}},
    url = {http://arxiv.org/abs/1606.07494},
    doi = {10.48550/arXiv.1606.07494},
    urldate = {2026-07-09},
    publisher = {arXiv},
    author = {Borsanyi, Sz and Fodor, Z. and Kampert, K. H. and Katz, S. D. and Kawanai, T. and Kovacs, T. G. and Mages, S. W. and Pasztor, A. and Pittler, F. and Redondo, J. and Ringwald, A. and Szabo, K. K.},
    month = jun,
    year = {2016},
    note = {arXiv:1606.07494 [hep-lat]},
}

@book{rajaraman_solitons_1984,
    address = {Amsterdam},
    edition = {2. print},
    title = {Solitons and instantons: an introduction to solitons and instantons in quantum field theory},
    isbn = {978-0-444-86229-7},
    shorttitle = {Solitons and instantons},
    language = {eng},
    publisher = {North-Holland},
    author = {Rajaraman, R.},
    year = {1984},
}

@article{blau_dynamics_1987,
    title = {Dynamics of false-vacuum bubbles},
    volume = {35},
    copyright = {http://link.aps.org/licenses/aps-default-license},
    issn = {0556-2821},
    url = {https://link.aps.org/doi/10.1103/PhysRevD.35.1747},
    doi = {10.1103/PhysRevD.35.1747},
    language = {en},
    number = {6},
    urldate = {2026-07-11},
    journal = {Physical Review D},
    author = {Blau, Steven K. and Guendelman, E. I. and Guth, Alan H.},
    month = mar,
    year = {1987},
    pages = {1747--1766},
}

@article{guven_covariant_1993,
    title = {Covariant perturbations of domain walls in curved spacetime},
    volume = {48},
    copyright = {http://link.aps.org/licenses/aps-default-license},
    issn = {0556-2821},
    url = {https://link.aps.org/doi/10.1103/PhysRevD.48.4604},
    doi = {10.1103/PhysRevD.48.4604},
    language = {en},
    number = {10},
    urldate = {2026-07-11},
    journal = {Physical Review D},
    author = {Guven, Jemal},
    month = nov,
    year = {1993},
    pages = {4604--4608},
}

@article{garriga_perturbations_1991,
    title = {Perturbations on domain walls and strings: {A} covariant theory},
    volume = {44},
    copyright = {http://link.aps.org/licenses/aps-default-license},
    issn = {0556-2821},
    shorttitle = {Perturbations on domain walls and strings},
    url = {https://link.aps.org/doi/10.1103/PhysRevD.44.1007},
    doi = {10.1103/PhysRevD.44.1007},
    language = {en},
    number = {4},
    urldate = {2026-07-11},
    journal = {Physical Review D},
    author = {Garriga, Jaume and Vilenkin, Alexander},
    month = aug,
    year = {1991},
    pages = {1007--1014},
}

@article{kain_are_2019,
    title = {Are gravitating magnetic monopoles stable?},
    volume = {100},
    issn = {2470-0010, 2470-0029},
    url = {http://arxiv.org/abs/1909.04205},
    doi = {10.1103/PhysRevD.100.063003},
    number = {6},
    urldate = {2026-07-18},
    journal = {Physical Review D},
    author = {Kain, Ben},
    month = sep,
    year = {2019},
    note = {arXiv:1909.04205 [gr-qc]},
    pages = {063003},
}

@article{monin_monopole_2005,
    title = {Monopole decay in the external electric field},
    volume = {2005},
    issn = {1029-8479},
    url = {http://arxiv.org/abs/hep-th/0509047},
    doi = {10.1088/1126-6708/2005/10/109},
    number = {10},
    urldate = {2026-07-23},
    journal = {Journal of High Energy Physics},
    author = {Monin, A. K.},
    month = oct,
    year = {2005},
    note = {arXiv:hep-th/0509047},
    pages = {109--109},
}

@misc{jeong_dissipation_2023,
    title = {Dissipation of axion energy via the {Schwinger} and {Witten} effects},
    url = {http://arxiv.org/abs/2309.16570},
    doi = {10.48550/arXiv.2309.16570},
    urldate = {2026-07-23},
    publisher = {arXiv},
    author = {Jeong, Kwang Sik and Nakagawa, Shota and Takahashi, Fuminobu and Yamada, Masaki},
    month = sep,
    year = {2023},
    note = {arXiv:2309.16570 [hep-ph]},
}

@article{donaldson_nahms_1984,
    title = {Nahm's equations and the classification of monopoles},
    volume = {96},
    issn = {0010-3616, 1432-0916},
    url = {https://projecteuclid.org/journals/communications-in-mathematical-physics/volume-96/issue-3/Nahms-equations-and-the-classification-of-monopoles/cmp/1103941858.full},
    language = {en},
    number = {3},
    urldate = {2026-07-23},
    journal = {Communications in Mathematical Physics},
    publisher = {Springer},
    author = {Donaldson, S. K.},
    month = jan,
    year = {1984},
    pages = {387--407},
}

@misc{harvey_magnetic_1996,
    title = {Magnetic {Monopoles}, {Duality}, and {Supersymmetry}},
    url = {http://arxiv.org/abs/hep-th/9603086},
    doi = {10.48550/arXiv.hep-th/9603086},
    urldate = {2026-07-23},
    publisher = {arXiv},
    author = {Harvey, Jeffrey A.},
    month = may,
    year = {1996},
    note = {arXiv:hep-th/9603086},
}

@article{baek_hidden_2014,
    title = {Hidden sector monopole, vector dark matter and dark radiation with {Higgs} portal},
    volume = {2014},
    issn = {1475-7516},
    url = {http://arxiv.org/abs/1311.1035},
    doi = {10.1088/1475-7516/2014/10/067},
    number = {10},
    urldate = {2026-07-27},
    journal = {Journal of Cosmology and Astroparticle Physics},
    author = {Baek, Seungwon and Ko, P. and Park, Wan-Il},
    month = oct,
    year = {2014},
    note = {arXiv:1311.1035 [hep-ph]},
    pages = {067--067},
}

@article{khoze_dark_2014,
    title = {Dark matter monopoles, vectors and photons},
    volume = {2014},
    issn = {1029-8479},
    url = {http://arxiv.org/abs/1406.2291},
    doi = {10.1007/JHEP10(2014)061},
    number = {10},
    urldate = {2026-07-27},
    journal = {Journal of High Energy Physics},
    author = {Khoze, Valentin V. and Ro, Gunnar},
    month = oct,
    year = {2014},
    note = {arXiv:1406.2291 [hep-ph]},
    pages = {61},
}

@article{gorsky_schwinger_2001,
    title = {Schwinger type processes via branes and their gravity duals},
    copyright = {Assumed arXiv.org perpetual, non-exclusive license to distribute this article for submissions made before January 2004},
    url = {https://arxiv.org/abs/hep-th/0110178},
    doi = {10.48550/ARXIV.HEP-TH/0110178},
    urldate = {2026-07-27},
    publisher = {arXiv},
    author = {Gorsky, A. S. and Saraikin, K. A. and Selivanov, K. G.},
    year = {2001},
    note = {Version Number: 3},
}

@article{hui_no-hair_2013,
    title = {No-{Hair} {Theorem} for the {Galileon}},
    volume = {110},
    copyright = {http://link.aps.org/licenses/aps-default-license},
    issn = {0031-9007, 1079-7114},
    url = {https://link.aps.org/doi/10.1103/PhysRevLett.110.241104},
    doi = {10.1103/PhysRevLett.110.241104},
    language = {en},
    number = {24},
    urldate = {2026-07-30},
    journal = {Physical Review Letters},
    author = {Hui, Lam and Nicolis, Alberto},
    month = jun,
    year = {2013},
    pages = {241104},
}

@misc{loladze_monopole-fermion_2024,
    title = {Monopole-{Fermion} {Scattering} and the {Solution} to the {Semiton}/{Unitarity} {Puzzle}},
    url = {http://arxiv.org/abs/2408.04577},
    doi = {10.48550/arXiv.2408.04577},
    urldate = {2026-08-07},
    publisher = {arXiv},
    author = {Loladze, Vazha and Okui, Takemichi},
    month = nov,
    year = {2024},
    note = {arXiv:2408.04577 [hep-th]},
}

@misc{hook_dyonic_2025,
    title = {Dyonic bound states},
    url = {http://arxiv.org/abs/2409.07549},
    doi = {10.48550/arXiv.2409.07549},
    urldate = {2026-08-07},
    publisher = {arXiv},
    author = {Hook, Anson and Ristow, Clayton},
    month = mar,
    year = {2025},
    note = {arXiv:2409.07549 [hep-th]},
}

\end{document}